\documentclass{ws-mplb}
\usepackage[super,sort,compress]{cite}
\usepackage{xcolor}
\usepackage[verbose,hypertexnames=false]{hyperref}
\hypersetup{colorlinks=false,allbordercolors=blue,pdfborderstyle={/S/U/W 1}}

\usepackage{graphicx} 
\usepackage{bm} 
\usepackage{amsmath} 
\usepackage{amssymb}  
\usepackage{booktabs}
\usepackage{tabularx}

\newcommand{\moire}{{moiré} }
\newcommand{\moirenospace}{{moiré}}

\begin{document}

\markboth{C.-H. Hsu and A. Ohorodnyk}{Coupled-wire descriptions of unconventional quantum states in twisted nanostructures}

%
\catchline{}{}{}{}{}
%

\title{Coupled-wire descriptions of unconventional quantum states in twisted nanostructures }

\author{Chen-Hsuan Hsu}

\address{Institute of Physics, Academia Sinica, Taipei 115201, Taiwan\\
Physics Division, National Center for Theoretical Sciences, Taipei 106319, Taiwan\\
chenhsuan@as.edu.tw}

\author{Anna Ohorodnyk}

\address{Institute of Physics, Academia Sinica, Taipei 115201, Taiwan\\
Department of Electronics and Electrical Engineering, National Yang Ming Chiao Tung University, Hsinchu 30010, Taiwan 
}

\maketitle


\begin{abstract}

Coupled-wire description has been developed as a powerful framework for providing bosonic descriptions of strongly correlated quantum matter, with early applications to systems such as the cuprates and the integer and fractional quantum Hall states. In this review, we discuss recent developments of coupled-wire description in nanoscale systems, where it emerges not only as a theoretical tool but also as a highly tunable physical platform.
In these nanoscale realizations, coupled-wire networks are formed by one-dimensional channels embedded in two-dimensional materials, most prominently in \moire and twisted structures. Such networks host a broad range of unconventional states of matter, including superconductivity, charge density waves, spin density waves, Mott insulating phases, Anderson insulating phases, quantum spin Hall states, quantum anomalous Hall states, and their fractionalized counterparts. The ability to electrically control interaction strength, confinement, and coupling between wires makes these systems qualitatively different from earlier realizations and allows continuous tuning between competing phases.
Notably, recent work has demonstrated that coupled-wire frameworks in \moire networks unify the trio of quantum Hall phenomena, encompassing quantum Hall, quantum spin Hall, and quantum anomalous Hall states, together with their fractional analogues. This development highlights coupled-wire networks in nanoscale materials as a versatile and experimentally relevant setting for exploring the interplay of topology, strong correlations, and low-dimensional physics. Throughout the article, the discussion is presented in non-technical terms, with minimal formalism, to make the underlying physical ideas accessible to a broad readership.

\end{abstract}

\keywords{coupled-wire construction; domain wall networks; moir\'e materials; sliding Luttinger liquids; electron-phonon coupling; 
superconductivity; 
topological phases; integer quantum anomalous Hall effect; fractional quantum anomalous Hall effect; fractional excitations; chiral edge modes; Anderson localization; spin helix order; electron-magnon coupling}


\section{Introduction}

One-dimensional  quantum electron systems  represent a unique platform in condensed matter physics. Unlike their higher-dimensional counterparts, interacting electrons in one dimension generically defy a Fermi-liquid description and instead realize collective modes governed by bosonic degrees of freedom.  
This paradigm was established through the pioneering works of Tomonaga~\cite{Tomonaga:1950} and Luttinger~\cite{Luttinger:1963}, and later formulated by Haldane into the modern framework of (Tomonaga-)Luttinger liquids~\cite{Haldane:1981}.  
A comprehensive and widely used account is provided in Giamarchi's formulation of interacting one-dimensional systems~\cite{Giamarchi:2003}.  
The success of bosonization in one dimension not only provides controlled access to strongly correlated physics, but also offers a flexible language to describe symmetry, topology, and collective excitations on equal footing.

Building on this foundation, the coupled-wire description emerged as a powerful route to access intrinsically higher-dimensional phases while retaining the analytical controllability of one-dimensional theories.\cite{Emery:2000,Vishwanath:2001,Mukhopadhyay:2001a,Mukhopadhyay:2001b}  
Early studies focused on constructing coupled-wire systems that realize sliding Luttinger liquids or smectic metal phases as candidate descriptions of the normal state in high-temperature superconductors, where pronounced deviations from Fermi-liquid behavior were observed.
Remarkably, by assembling arrays or networks of interacting quantum wires, one can engineer a wide variety of two-dimensional and even three-dimensional topological phases.\cite{Kane:2002,Klinovaja:2013b,Klinovaja:2014,Klinovaja:2014b,Neupert:2014,Sagi:2014,Teo:2014,Klinovaja:2015,Santos:2015,Fuji:2019,Imamura:2019,Meng:2020,Fuji:2023}
Early and influential examples include the construction of fractional and integer quantum Hall  states, where bulk topological order and protected edge modes arise from interwire couplings.\cite{Kane:2002}  
Among others, subsequent developments extended this framework to quantum spin Hall insulators,\cite{Klinovaja:2014} symmetry-protected topological states,\cite{Neupert:2014} non-Abelian topological orders,\cite{Sagi:2014,Teo:2014}  skyrmion-driven topological phases,\cite{Klinovaja:2015} and fracton order.\cite{Fuji:2023}
In this perspective, topology is encoded in the locking of collective bosonic fields, beyond the single-particle band theory.
 
More recently, this coupled-wire description has found renewed relevance in \moire materials formed by twisted van der Waals heterostructures.\cite{Andrei:2020,Balents:2020,Andrei:2021,Kennes:2021} 
The realization of magic-angle twisted bilayer graphene highlighted the dramatic role of \moire superlattices in generating flat bands and strong correlation effects.\cite{Cao:2018a,Cao:2018b} However, beyond flat-band formation, twisted layered structures also provide an alternative and conceptually distinct route to strong correlations through dimensional reduction. In particular, twisted multilayers tend to form locally preferred stacking domains separated by narrow domain walls.\cite{San-Jose:2013,Alden:2013,Carr:2018,Covaci:2018,Efimkin:2018,Huang:2018,Ramires:2018,Rickhaus:2018,Carr:2019,Xu:2019,Yoo:2019,Fleischmann:2020,Hou:2020,Tsim:2020,Walet:2020,Verbakel:2021,Mondal:2023,Wittig:2023,Wang:2024} 
A particularly important setting is small-angle twisted bilayer graphene under an interlayer bias, where the low-energy electronic states in neighboring AB- and BA-stacking regions can be described by massive Dirac cones, with the sign of the Dirac mass determined by the local stacking configuration. As a result, the boundaries between these regions, namely the domain walls, support gapless one-dimensional modes.\cite{San-Jose:2013,Carr:2018,Efimkin:2018} 
Along these domain walls, electronic states are spatially confined and effectively one-dimensional, forming a natural realization of quantum wire networks. As these domain walls proliferate across the \moire pattern, this provides a natural setting in which a two-dimensional \moire system can be recast as a network of coupled one-dimensional channels.

Importantly, the realization of  quantum wire networks in twisted nanostructures does not rely on fine-tuning to a magic angle nor on the formation of global flat bands. Instead, correlation effects set in as a consequence of the confinement of carriers into one-dimensional channels, where interactions are intrinsically enhanced and bosonization-based descriptions become appropriate.\cite{Wu:2019,Chou:2019,Chen:2020,Konig:2020,Lee:2021,Hsu:2023,Wang:2024,Chang:2025}
From this viewpoint, \moire domain wall networks provide a highly tunable, nanoscale platform that naturally bridges traditional coupled-wire description and experimentally accessible materials.  
They enable the realization of diverse correlated quantum phenomena, including quantum anomalous Hall states, superconductivity, charge density waves, disorder-induced Anderson insulating and spin helix phases, within a single, geometrically controlled framework.

In this review, we discuss the application of the correlated networks to two-dimensional nanoscale materials, where interacting one-dimensional channels emerge naturally from spatial confinement and structural modulation.  
We focus on twisted bilayer   systems, where \moire domain wall networks effectively realize arrays of interacting quantum wires without relying on magic-angle tuning or flat bands. 
We discuss how the bosonization description has been applied to establish various unconventional quantum states.  
Throughout the review, we emphasize bosonic descriptions and interaction-driven phenomena relevant to experimentally accessible \moire materials.

 The remainder of this review is organized as follows.  
In Sec.~\ref{Sec:network}, we introduce the coupled-wire description of interacting electrons confined to domain wall modes in nanoscale twisted structures.  
The quadratic Hamiltonian, which defines the correlated domain wall network fixed point, is introduced in Sec.~\ref{SubSec:quadratic}, followed by an analysis of its instabilities toward various electronic phases in Sec.~\ref{SubSec:instability}.  
Unconventional quantum states driven by non-quadratic terms are addressed in Sec.~\ref{Sec:BeyondQuadratic}.  
We discuss Anderson localization of the domain wall network in 
Sec.~\ref{SubSec:AndersonLocalization}. 
The density wave and superconducting orders arising from electron-electron interactions are discussed in Sec.~\ref{SubSec:ee-scattering}.
We show how \moire potential can give rise to quantum anomalous Hall states and gapless chiral modes in 
Sec.~\ref{SubSec:moire-umklapp}. 
In Sec.~\ref{Sec:BeyondTBG}, we extend the coupled-wire perspective beyond purely electronic systems and twisted bilayer graphene.  
Section~\ref{SubSec:2Dhelix} focuses on systems with coexisting localized magnetic moments and interacting itinerant carriers, where a two-dimensional spin helix emerges at low temperatures.  
Alternative platforms based on different twisted structures or materials are discussed in Sec.~\ref{SubSec:Others}.  
Finally, we conclude with an outlook in Sec.~\ref{Sec:Outlook}.

\section{Correlated domain wall network}
\label{Sec:network}

Twisted layered structures generically develop spatially varying local stacking configurations. 
In twisted bilayer graphene, the application of a perpendicular electric displacement field opens local spectral gaps in the AB- and BA-stacking regions with opposite signs of a valley Hall mass, while leaving one-dimensional domain walls that host gapless low-energy modes.\cite{San-Jose:2013,Alden:2013,Carr:2018,Covaci:2018,Efimkin:2018,Huang:2018,Ramires:2018,Rickhaus:2018,Carr:2019,Xu:2019,Yoo:2019,Fleischmann:2020,Hou:2020,Tsim:2020,Walet:2020,Verbakel:2021,Mondal:2023,Wittig:2023,Wang:2024}   
The central point for this review is that this platform realizes interacting one-dimensional channels embedded in two dimensions without requiring magic-angle fine-tuning or globally flat bands. 
Instead, correlation effects are promoted by the geometric confinement to one dimension and the resulting enhancement of interaction effects characteristic of Tomonaga-Luttinger liquids.

\begin{figure}[th]
\centerline{
\includegraphics[width=0.55\linewidth]{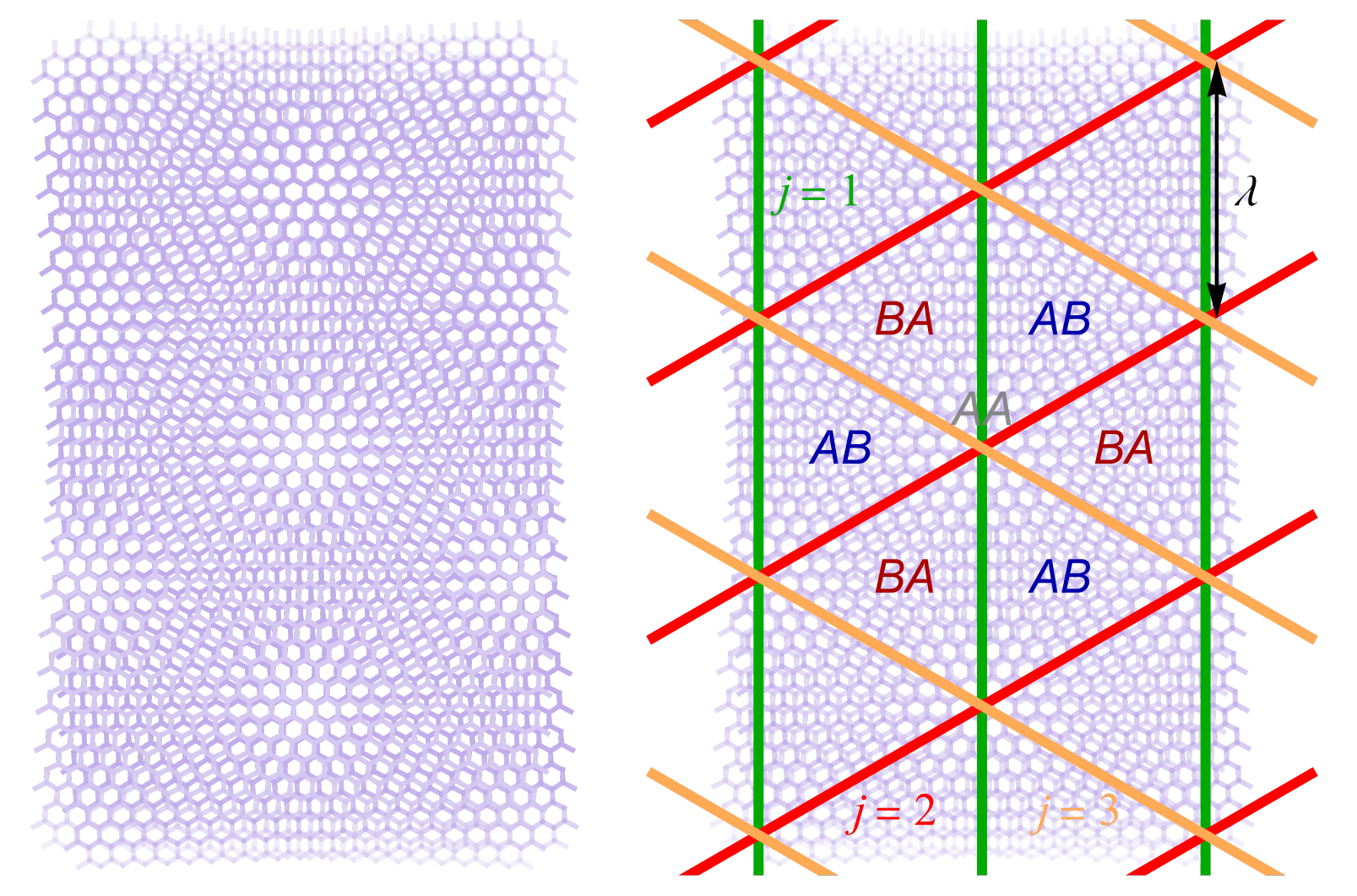}
\includegraphics[width=0.45\linewidth]{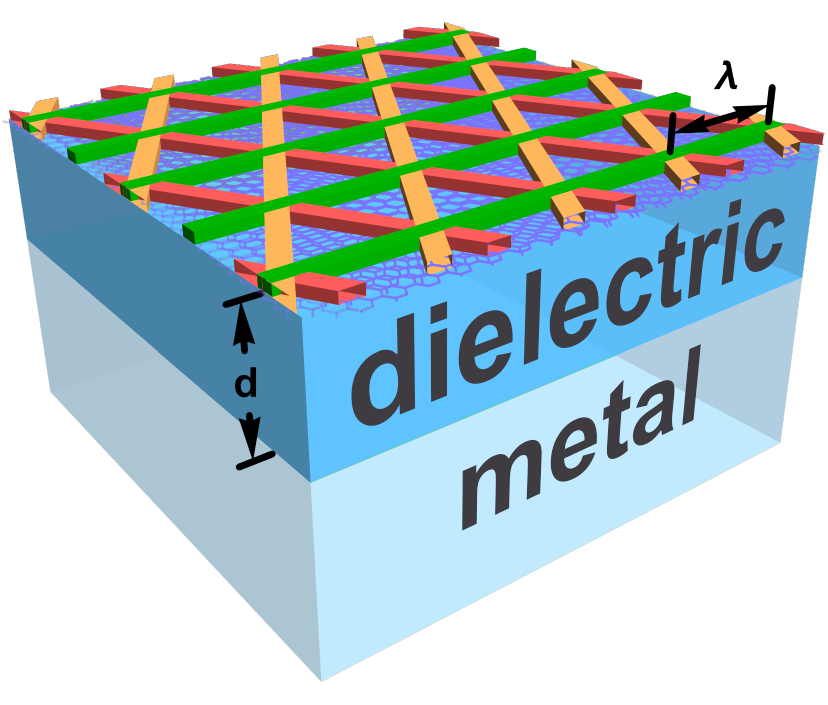}
}
\caption{Left: Schematic illustration of the \moire pattern in twisted bilayer graphene (purple). 
Spatially varying stacking configurations generate AB- and BA-stacking domains, with the domain walls (green, red, and orange) forming a triangular network.  
Right: Illustration of the device nanostructure, consisting of twisted graphene layers (purple), a dielectric spacer (blue), and a metallic gate (light blue).
}
\label{Fig:TBG_dw}
\end{figure}

Because the domain walls appear in three sets rotated by $120^{\circ}$, the resulting structure naturally forms a two-dimensional triangular quantum network; see Fig.~\ref{Fig:TBG_dw} for illustration. 
A key feature of this platform is its high degree of electrical tunability. 
As illustrated in Fig.~\ref{Fig:TBG_dw}, one can build a device that integrates the system with a nearby metallic gate. 
Device control parameters, such as the interlayer bias $V_d$ and electrostatic screening, governed by the gate distance $d$ and dielectric environment, can strongly reshape the spatial profiles of the domain wall wave functions. 
As a consequence, the domain wall Fermi velocity, the effective bandwidth, and the strength of electron-electron interactions within and between domain walls can all be tuned {\it in situ}~\cite{Wang:2024}. 
This enables systematic navigation between weakly coupled quantum wires and strongly correlated coupled-wire regimes while keeping the network geometry essentially fixed.

From a broader perspective, this domain wall network provides a complementary route to correlated physics in twisted bilayer graphene that is largely insensitive to the precise stacking configuration responsible for magic-angle flat bands. 
Electrical control alone allows the network to access distinct correlated regimes, including density-wave order, pairing instabilities, and Mott or Anderson insulating behaviors.
Hence, twisted bilayer graphene can host a rich landscape of correlated phenomena across a wide range of twist angles and device configurations, provided that the low-energy physics is dominated by domain wall modes rather than bulk \moire bands.

\subsection{Quadratic fixed point: electrically tunable correlated network }

\label{SubSec:quadratic}

\begin{figure}[h]
\centerline{
\includegraphics[width=0.3\linewidth]{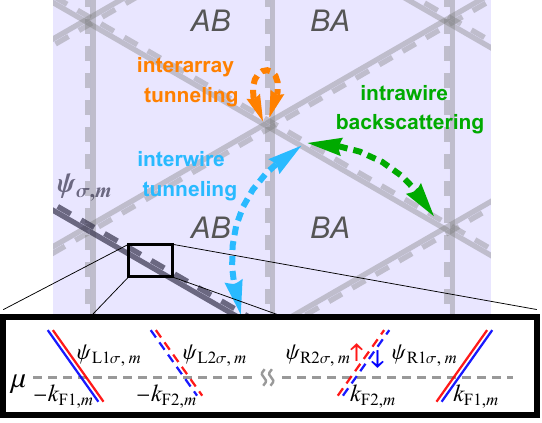}
\includegraphics[width=0.26\linewidth]{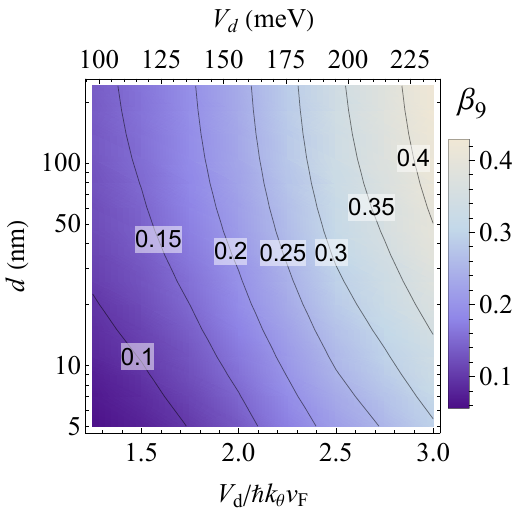}
\includegraphics[width=0.36\linewidth]{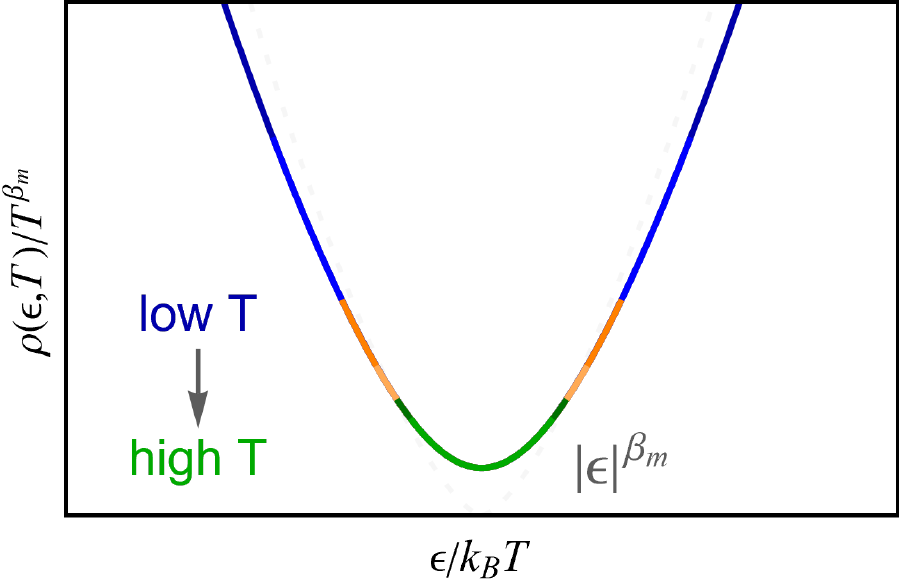}
} 
\caption{Left: Schematic illustration of the domain wall network and its low-energy spectrum. In the enlarged spectrum of a domain wall labeled by $m$, there are two energy band branches $\delta=1$ (solid) and $\delta=2$ (dashed) with spin up (red) and spin down (blue) states. The illustrated scattering and tunneling processes include intrawire backscattering (green), interwire tunneling (light blue), and interarray tunneling (orange). 
Middle: Scaling exponent ($\beta_{9}$) of the local density of states defined in Eq.~\eqref{Eq:DOS}. This plot is reproduced from Ref.~\citenum{Wang:2024}. 
Right: Universal collapse of the density-of-states curves measured at different temperatures.  
}
\label{Fig:scattering-network} 
\end{figure}

We now build up the network description of the interacting domain wall modes, which bridges continuum single-particle models and the many-body bosonized theory. 
Concretely, one introduces the fermion fields
$\psi_{\sigma, m} = \sum_{\ell,\delta} \psi_{\ell \delta \sigma, m}$
for each domain wall shown in Fig.~\ref{Fig:TBG_dw}, together with its bosonized representation, 
\begin{align} 
\psi_{\ell \delta \sigma ,m} (x) &=  \frac{ U_{\ell \delta \sigma, m}}{\sqrt{2\pi a}} 
e^{i \ell k_{F \delta, m} x} 
\exp \bigg\{\frac{i}{2}\Big[- \ell  (\phi_{cS, m} + \delta \phi_{cA, m} )  
- \ell  \sigma (\phi_{sS,m} + \delta \phi_{sA, m})
\nonumber \\
& \quad    
+ (\theta_{cS, m} + \delta \theta_{cA, m} )  
+ \sigma (\theta_{sS, m} + \delta \theta_{sA, m}) \Big] \bigg\}, 
\label{Eq:bosonization}
\end{align}
where $m$ indexes the domain walls parallel to each other, $\ell \in \{ R \equiv + , L \equiv - \}$ denotes the propagation direction within a domain wall, $\sigma \in \{\uparrow \equiv + , \downarrow \equiv - \}$ is the spin index, $x$ is the coordinate along the domain wall, $k_{F\delta,m}$ is the Fermi wave vector, and $\delta \in \{1 \equiv +, 2 \equiv -\}$ labels the two low-energy branches in the domain wall spectrum; see Fig.~\ref{Fig:scattering-network} for the illustration.
In the above, we introduce the short-distance cutoff $a$. 

In the diagonalized basis, the original spin index is reorganized into 
$\nu \in \{c,s\}$, labeling the charge and spin sectors, while the branch index is mapped to 
$P \in \{S,A\}$, corresponding to symmetric and antisymmetric combinations of the energy band branches. 
In this representation, the bosonic fields $\phi_{\nu P,m}$ and $\partial_x \theta_{\nu P,m}$ form a pair of canonically conjugate fields.

At the quadratic level, the system realizes a network generalization of a Tomonaga-Luttinger liquid with spin and valley degrees of freedom: each domain wall is effectively described as a carbon nanotube,\cite{Kane:1997} while the network structure enters through interactions between parallel domain walls. 
The bosonized formulation is organized into charge and spin sectors and, owing to the presence of multiple energy band branches in the domain wall spectrum, into symmetric and antisymmetric combinations between branches. 
The resulting effective Hamiltonian takes the quadratic form,
\begin{subequations}
\label{Eq:H0_eqset}
\begin{align}
H_{\mathrm{ee}} & = H_{cS} + H_{cA} + H_{sS}+  H_{sA}  , 
\label{Eq:H0_original} 
\end{align}
with one interacting sector, 
\begin{align}
H_{cS} &= \sum_{m,n} \int \frac{ dx}{2\pi} 
\bigg[ U_{\phi_{cS},n} 
\left( \partial_x \phi_{cS, m} \right) 
\left( \partial_x \phi_{cS, m + n } \right)  
+ U_{\theta_{cS},n} 
\left( \partial_x \theta_{cS, m} \right) 
\left( \partial_x \theta_{cS, m+n} \right) \bigg],  
\label{Eq:H0cS_original} 
\end{align}
where $U_{\phi_{cP},n}$ and $U_{\theta_{cP},n}$ quantify, respectively, the density-density and current-current interaction strengths between the $n$th-nearest-neighbor parallel domain walls. 
 The other sectors ($\nu P \neq cS$) are effectively noninteracting\cite{Wang:2024}
\begin{align}
H_{\nu P } &= \sum_{m} \int \frac{\hbar dx}{2\pi} 
\bigg[ \frac{u_{ \nu P}}{K_{ \nu P}} 
\left( \partial_x \phi_{\nu P, m} \right)^2  
+ u_{\nu P} K_{\nu P} 
\left( \partial_x \theta_{\nu P, m} \right)^2 \bigg], 
\label{Eq:H0s_original}
\end{align}
\end{subequations}
where  we have $u_{\nu P} = v_{\rm dw} / K_{\nu P}$ and the interaction parameter $K_{cA} = K_{sS} = K_{sA} = 1$. 

Starting from the continuum description,\cite{Bistritzer:2011,Nam:2017,Koshino:2018,Efimkin:2018,Carr:2019,Tarnopolsky:2019,Koshino:2020,Nakatsuji:2023}  
one identifies the domain wall channels and their low-energy branches, including their velocities, Fermi momenta, and, crucially, their real-space charge density profiles. These microscopic ingredients determine the forward-scattering part of the interacting problem: density-density couplings \emph{within} a given domain wall (intrawire) and \emph{between} parallel domain walls (interwire). Ref.~\citenum{Wang:2024} details this procedure, including how to extract domain wall density profiles and convert them into effective interaction parameters for realistic device geometries.
Another consideration is perturbations at intersections, as shown in Fig.~\ref{Fig:scattering-network}. 
Scattering processes involving crossing domain walls can occur at the network nodes. However, since such terms typically enter the effective action without a spatial integral, they are less relevant in the renormalization-group sense than processes confined to a single domain wall or between parallel domain walls~\cite{Chen:2020,Hsu:2023}. 
This hierarchy motivates focusing first on the quasi-one-dimensional building blocks of the network and their couplings along parallel domain walls, as described by the quadratic Hamiltonian in Eq.~\eqref{Eq:H0_eqset}.

Within the bosonized framework, correlation effects arising from electron-electron interactions can be analyzed by examining the scaling behavior of correlation functions. 
For example, the correlators in the charge-symmetric sector  take the form,
\begin{subequations}
\begin{align}
    \left\langle  
e^{ - \sqrt{2}\, i \phi_{cS,m} (x, \tau) } 
e^{ \sqrt{2}\, i \phi_{cS,m} (0) } 
\right\rangle_{\rm ee} 
\propto &
\left| \frac{a}{ \sqrt{x^2 + (u |\tau| + a)^2 } } \right|^{\Delta_{\phi_{cS,m}} },
\\ 
\left\langle  
e^{ - \sqrt{2}\, i \theta_{cS,m} (x, \tau) } 
e^{ \sqrt{2}\, i \theta_{cS,m} (0) } 
\right\rangle_{\rm ee} 
\propto &
\left| \frac{a}{ \sqrt{x^2 + (u |\tau| + a)^2 } } \right|^{\Delta_{\theta_{cS,m}} } ,
\end{align}
\end{subequations}
with the imaginary time $\tau$. 
For noninteracting electrons, the scaling exponents reduce to unity, whereas stronger interactions lead to larger deviations from unity.
Therefore, only the parameter of the charge-symmetric sector labeled by $\nu =$ c and $P=$ S will enter the exponents explicitly.   

A key feature of coupled wires in \moire domain walls is that these scaling exponents are electrically tunable, as both the interlayer bias $V_d$ and the electrostatic screening length $d$ influence the interaction strength; see Fig.~\ref{Fig:TBG_dw} for the device geometry which can be used to control $d$. 
Conceptually, the roles of these knobs are straightforward: increasing $V_d$ enhances the transverse confinement of the domain wall modes and thereby strengthens effective density-density interactions, while decreasing $d$ enhances screening. 
Remarkably, correlation strengths and dominant low-energy tendencies can also be adjusted \emph{in situ} without altering the underlying network geometry.

Even at this quadratic level, the framework already provides several directly testable consequences. In particular, experimentally accessible probes such as scanning tunneling spectroscopy can be systematically connected onto the underlying correlation functions and their associated scaling dimensions, providing a direct connection between device control parameters and low-energy observables.

An example is the system's spectroscopic features. Specifically, when probing a domain wall, the local density of states exhibits the familiar one-dimensional interaction fingerprint: power-law suppression near the Fermi level,  which can be expressed as a function of energy $\epsilon$ and temperature $T$~\cite{Wang:2024},
\begin{equation}
    \rho_{{\rm dos},m} (\epsilon,T) \propto  T^{\beta_{m} } \cosh \left( \frac{ \epsilon }{2 k_{\rm B} T}  \right)   \left| \Gamma \left( \frac{1+\beta_{m} }{2} +  \frac{i \epsilon }{2 \pi k_{\rm B} T } \right) \right|^2,  
 \label{Eq:DOS} 
 \end{equation}
with an interaction-dependent exponent $\beta_{m} $ determined by the  interaction strength of the entire network, rather than that of a single isolated wire.  
Two features are illustrated in Fig.~\ref{Fig:scattering-network}.
First, the formula for the density of states exhibits a universal scaling behavior, with data points measured at different temperatures collapsing onto one single curve upon rescaling.
Second, the scaling exponent is electrically tunable with $V_d$ and $d$, providing an experimental knob to quantify correlations.\cite{Wang:2024} 
In mesoscopic devices, the exponent may vary  across the sample (thus depending on the index $m$)  when translational invariance perpendicular to the domain walls is not assumed. 
 
Beyond spectroscopy, correlated signatures also appear in microscopic scattering processes related to charge transport.
At this level, three processes occur throughout the network, as illustrated in Fig.~\ref{Fig:scattering-network}: 
(i) backscattering from impurities within domain walls, (ii) tunneling between crossing domain walls at an intersection, and (iii) tunneling between {parallel} domain walls. Each process is governed by a scaling dimension fixed by the quadratic  theory in Eq.~\eqref{Eq:H0_eqset} 
and therefore produces characteristic power-law dependencies on temperature and bias. 
The exponents are again set by the forward-scattering terms in the quadratic  theory and thus tunable by $V_d$ and $d$, whereas the overall amplitudes depend on nonuniversal parameters, such as impurity strength and microscopic wave-function overlap.

\subsection{Instability of the network toward various electronic phases}
\label{SubSec:instability}

Electronic phases of the domain wall network can be systematically explored through their correlation functions.  
A key theoretical advantage of formulating the network as a coupled Tomonaga-Luttinger liquid is that it provides a compact and transparent criterion for identifying the dominant low-energy tendencies of the system.  
In one dimension, correlation functions decay as power laws,\cite{Haldane:1981,Giamarchi:2003} with exponents that depend sensitively on interaction strength.  
The correlator(s) with the slowest decay therefore signals the leading instability as the system is cooled or probed at long length scales.

In practice, one selects a set of candidate operators and compares their associated correlation exponents.  
For density-wave tendencies, a natural choice is the $2k_F$ components of the charge and spin densities, corresponding to charge-density-wave and spin-density-wave channels.  
For superconducting tendencies, one instead examines pairing operators describing singlet and triplet superconducting correlations.  

Because each domain wall supports multiple low-energy branches in the present setting, both intrabranch and interbranch density-wave operators are allowed,\cite{Wang:2024} closely analogous to the situation in metallic carbon nanotubes.\cite{Egger:1997,Egger:1998,Hsu:2015} 
Pairing operators with zero total momentum are naturally intrabranch, whereas interbranch pairing carries finite momentum and is therefore typically subleading when diagnosing the leading instability of the network.

\subsubsection{Purely electronic system: density-wave dominance}

For a purely electronic coupled-wire system, the candidate correlation functions between $n$th-nearest neighbor wires take the standard power-law form, 
\begin{equation}
\langle O_{m}^\dagger(x,\tau) O_{m+n} (0)\rangle \propto \left|\frac{a}{ \sqrt{x^2 + u^2 \tau^2 } }\right|^{\Delta_n},
\end{equation}
where $O_m$ represents the operator on wire $m$ that characterizes the density wave or pairing, and the scaling exponent $\Delta_n$ can be expressed in terms of the interaction-dependent parameters introduced in Eq.~\eqref{Eq:H0_eqset}.  
The dominant low-energy tendency is identified by the smallest exponent $\Delta_n$ among those channels satisfying the instability condition $\Delta_n < 2$.

In the experimentally relevant regime where repulsive interactions are enhanced by one-dimensional confinement, density-wave correlations are typically favored over pairing correlations.  
As a result, the leading instability of a purely electronic network tends to be toward charge-density-wave or spin-density-wave order.  
When spin-rotation symmetry is preserved and the spin sector remains effectively noninteracting, the charge and spin density-wave channels can have the same scaling exponents, so the eventual selection between them may depend on weaker symmetry-breaking perturbations beyond the quadratic fixed point.

A specific feature in coupled-wire systems is worth emphasizing.  
Because \moire domain wall networks in mesoscopic devices do not necessarily exhibit exact translational invariance perpendicular to the domain walls, the interaction parameters and hence the scaling exponents can vary weakly with the domain wall index.  
This naturally leads to a crossover scenario in which different domain walls become unstable at slightly different energy or temperature scales, rather than undergoing a  sharp global transition.

\subsubsection{Phonon-enhanced pairing and a tunable competition}
\label{Sec:Phonon-in-TBG}

Electron-phonon coupling can qualitatively modify the balance between density-wave and superconducting tendencies.  
Although microscopic models for electron-phonon coupling in domain wall networks are still under development, it is instructive to consider an effective description in which longitudinal acoustic phonons couple to the low-energy charge density of each channel,\cite{Wentzel:1951,Bardeen:1951,Engelsberg:1964,Loss:1994,Martin:1995,Hsu:2024} 
in close analogy with standard one-dimensional electron-phonon problems.  
In the bosonized formulation, this coupling acts primarily on the charge-symmetric sector and breaks the simple duality between the $\phi$ and $\theta$ fields that holds in a purely electronic Tomonaga-Luttinger liquid,\cite{Furusaki:1993}  
As a result, phonons renormalize the correlation exponents in opposite directions for density-wave and pairing channels: density-wave correlations are suppressed, while pairing correlations are enhanced.  
This behavior reflects the standard mechanism by which phonons mediate an effective attractive interaction and promote superconducting tendencies in low-dimensional systems.\cite{Wentzel:1951,Bardeen:1951,Engelsberg:1964,Loss:1994,Martin:1995,Hsu:2024}  

To make this explicit, we consider electron-phonon coupling in the one-dimensional channels of the coupled-wire network and introduce a minimal model for longitudinal acoustic phonons confined to a domain wall,
\begin{align}
H_{\rm ph}  = \sum_{m} \frac{1}{2 \rho_{\rm a}} \int dx 
\left[
\left(\Pi_{{\rm ph},m}\right)^2 
+ \rho_{\rm a}^2 c_{\rm ph}^2 
\left(\partial_x d_{{\rm ph},m}\right)^2
\right],
\label{Eq:H_ph}
\end{align}
where $\rho_{\rm a}$ denotes the effective mass density distributed along the domain wall, $c_{\rm ph}$ is the phonon velocity, $d_{{\rm ph},m}$ is the lattice displacement field, and $\Pi_{{\rm ph},m}$ is its conjugate momentum.  
The deformation potential induced by phonons couples to the electronic density within each domain wall, hence leading to the Holstein type of electron-phonon coupling.  
In the bosonized description, this coupling is expressed as
\begin{equation}
H_{{\rm ep}} = \sum_{m} g_{{\rm ep}} \int dx \,
\left(\partial_x \phi_{cS,m}\right)
\left(\partial_x d_{{\rm ph},m}\right),
\label{Eq:H_ep}
\end{equation}
with $g_{\rm ep}$ the effective electron-phonon coupling strength. 

The hybridization between the electronic and phononic degrees of freedom produces two collective modes with renormalized velocities.\cite{Wang:2024}  
Importantly, this formulation goes beyond a perturbative treatment in $g_{\rm ep}$ and captures the full hybridization of electronic and phononic modes.  
Upon integrating out the phonon fields, the scaling dimensions of electronic correlation functions are modified, with the resulting exponents encoding nonperturbative effects of the electron-phonon coupling.  
Consequently, operators involving $\phi_{cS}$ become less renormalization-group relevant, while those involving $\theta_{cS}$ become more relevant.  
Physically, this reflects the emergence of an effective phonon-mediated attraction in the charge-symmetric sector.

\begin{figure}[th]
\centerline{\includegraphics[width=\linewidth]{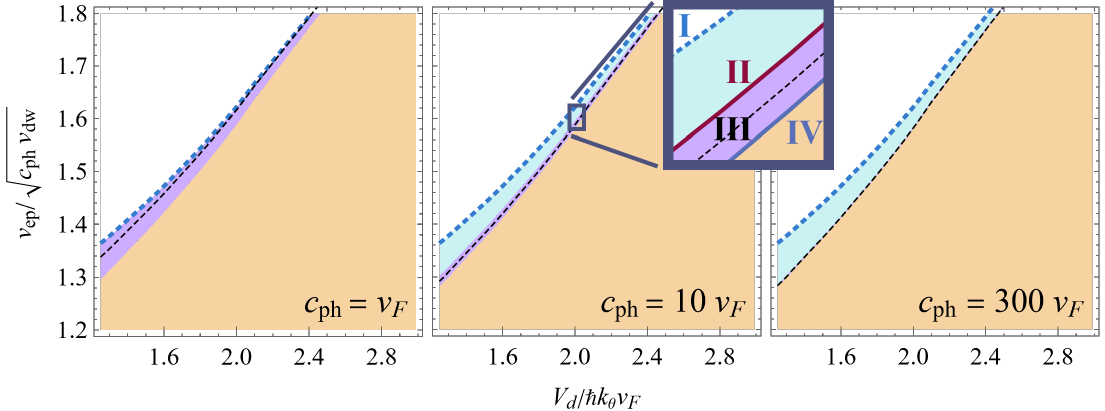}}
\caption{Phase diagram of correlated coupled wires in twisted bilayer graphene as a function of interlayer bias ($V_d$) and electron-phonon coupling. 
The phases include a gapless correlated network phase (purple), a density-wave-dominated regime (orange), a phonon-assisted superconducting regime (blue), and an electron-phonon-coupled liquid (white). 
The figure is reproduced from Ref.~\citenum{Wang:2024}.
}
\label{Fig:PhaseDiagram_network}
\end{figure}

Taking these modified exponents into account results in a phase diagram in which density-wave-dominated regimes compete with phonon-assisted superconducting regimes, as summarized in Fig.~\ref{Fig:PhaseDiagram_network}.  
From the network perspective, the qualitative outcome is a tunable competition among three regimes:  
(i) a gapless correlated network phase described by the quadratic fixed point;  
(ii) a density-wave-dominated regime favored by repulsive electron-electron interactions; and  
(iii) a phonon-assisted superconducting regime realized at a sufficiently strong electron-phonon coupling and a weaker electronic repulsion.

In addition, electron-phonon coupling introduces a conceptually important constraint, the Wentzel-Bardeen singularity, familiar from one-dimensional electron-phonon systems.  
At this singularity, a hybridized collective mode softens and its velocity vanishes, signaling a breakdown of the effective low-energy description and the emergence of a strongly coupled electron-phonon liquid~\cite{Wentzel:1951,Bardeen:1951,Engelsberg:1964,Loss:1994,Martin:1995}.  

A distinctive feature of domain wall networks is that the location of the Wentzel-Bardeen singularity is not fixed by the material parameters alone.  
Because the hybridization depends on both phonon properties and interaction-renormalized electronic velocities, the proximity to this singularity is electrically tunable.  
This provides a highly tunable route to phonon-enhanced superconductivity in a nanoscale two-dimensional setting, a level of control that is far less accessible in conventional one-dimensional systems.

Overall, the correlation-function viewpoint provides a practical map from tunable device parameters to low-energy phases.  
It distinguishes the role of the quadratic fixed point, which determines the scaling dimensions, from that of weaker perturbations, including junction processes, backscattering and symmetry-allowed couplings, which ultimately govern the leading one-dimensional instabilities.

In the sections below, we use this framework to show how \moire domain wall networks offer a tunable route to correlated one-dimensional physics within a two-dimensional setting. We further demonstrate how this perspective unifies spectroscopic, transport, and ordering phenomena into a single
description.

\section{Electronic phases induced by non-quadratic perturbations}
\label{Sec:BeyondQuadratic}

While the quadratic Hamiltonian provides a controlled starting point and fixes the scaling dimensions of operators, it does not by itself determine the ultimate low-energy fate of the system.  
Going beyond the quadratic terms, one can incorporate interaction-generated scattering processes such as backscattering and pairing.  
These perturbations are non-quadratic in the bosonic fields and therefore cannot be diagonalized exactly.  
Their effects are instead analyzed within a perturbative renormalization-group framework, where the scaling dimensions set by the quadratic fixed point determine whether a given process is relevant, marginal, or irrelevant at low energies.

Within this framework, the domain wall network offers a natural setting to investigate interaction-driven insulating phases, including Mott insulating phase arising from scattering processes  of interactions and Anderson insulating phase driven by disorder.

\subsection{Anderson localization of the domain wall network}
\label{SubSec:AndersonLocalization}

Given the quasi-one-dimensional nature of the domain wall modes, disorder constitutes a natural and often unavoidable class of perturbations beyond the quadratic fixed point.\cite{Abrahams:1979, Abrahams:2010}   
At low energies, such perturbations can be amplified by interactions and may ultimately drive the system toward Anderson localization.  

A convenient description models disorder on the $m$th domain wall 
by a random potential $V_{{\rm dis},m}(x)$ coupled to the electron charge density.
In the fermionic language, this random potential induces elastic backscattering within each domain wall,
\begin{equation}
H_{{\rm dis},m} =
\sum_{\delta \delta' \sigma} \int dx \;
V_{{\rm dis},m} (x)
 \psi_{R \delta \sigma,m} ^\dagger (x) 
\psi_{L \delta' \sigma,m}(x)
+ \text{H.c.} ,
\label{Eq:H_dis}
\end{equation}
which can be treated using standard disorder-averaged  techniques,\cite{Giamarchi:2003}
leading to a disorder strength subject to the same renormalization-group flow as in an interacting one-dimensional wire.\cite{Giamarchi:1988,Furusaki:1993,Giamarchi:2003}
In practice, the effective disorder acting on domain wall modes may be significantly weaker than that in bulk two-dimensional graphene. This suppression arises from suppressed intervalley scattering, closely analogous to the protection of helical channels in quantum spin Hall systems.\cite{Hsu:2017,Hsu:2018b,Hsu:2021}

Notably, the disorder-induced backscattering is relevant even for weakly attractive interactions, emphasizing that the tendency toward Anderson localization is governed primarily by interaction-renormalized exponents rather than by the bare disorder strength alone.  
Consequently, at sufficiently long length scales or low temperatures, a domain wall network with finite disorder flows toward a localized regime.

\begin{figure}[th]
\centerline{
\includegraphics[width=0.4\linewidth]{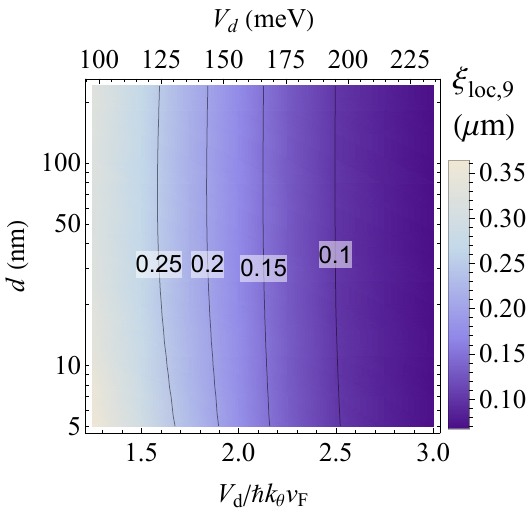}
\hspace{0.05\linewidth}
\includegraphics[width=0.4\linewidth]{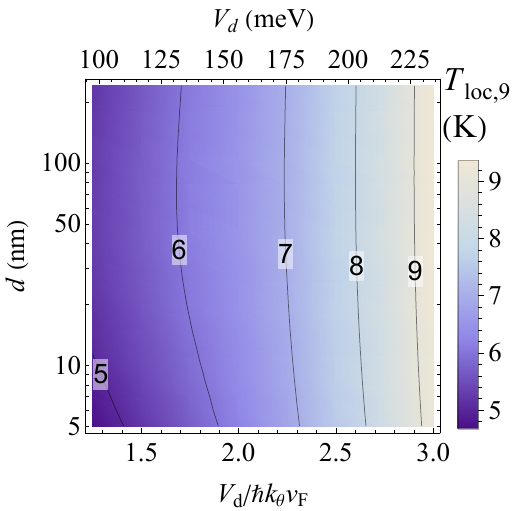}
} 
\caption{Localization length ($\xi_{\rm loc}$) and temperature ($T_{\rm loc}$) of the domain wall network as a function of interlayer bias ($V_d$) and screening length ($d$). 
The figures are reproduced from Ref.~\citenum{Wang:2024}. 
}
\label{Fig:Localization} 
\end{figure}

It is useful to summarize the consequences of this flow in terms of two physical scales: a localization length $\xi_{{\rm loc}}$ and a corresponding localization temperature $T_{{\rm loc}}$.  
Up to nonuniversal prefactors, both quantities can be extracted directly from the renormalization-group equations,\cite{Wang:2024}
and representative results are shown in Fig.~\ref{Fig:Localization}.  
The resulting localization properties are controllable. Increasing the interlayer bias typically enhances electron-electron interactions and thus favors localization, whereas stronger electrostatic screening suppresses long-range interwire couplings and can partially counteract this trend.

Estimates based on experimentally reasonable disorder strengths indicate that, for sufficiently long domain walls and at low temperatures, the network can cross over into an Anderson insulating regime.  
Importantly, this implies that localization can be tuned and monitored within a single device and a single disorder realization, for example by varying interlayer bias or screening during a single cooldown.

From the coupled-wire perspective, Anderson localization is therefore not merely a materials-dependent limitation, but an intrinsic competing tendency of correlated domain wall networks.  
As device sizes are scaled up or temperatures are lowered, localization effects become increasingly pronounced.  
At the same time, $\xi_{{\rm loc}}$ and $T_{{\rm loc}}$ remain electrically tunable within a given sample, providing a controlled platform to study interaction-driven metal-insulator crossovers in a nanoscale setting.

\subsection{Scattering from electron-electron interactions}
\label{SubSec:ee-scattering}

In addition to disorder-induced backscattering, electron-electron interactions themselves can generate backscattering processes, even in nominally clean systems.  In one-dimensional and quasi-one-dimensional settings, such interaction-driven backscattering terms are well known to destabilize the Tomonaga-Luttinger liquid fixed point and to drive the system toward a variety of ordered or gapped phases, depending on symmetry and commensurability.
In the context of twisted bilayer graphene, extending this logic beyond single-wire models was pioneered in a series of early network-based studies.  
Initial analyses of twisted systems~\cite{Wu:2019,Chou:2019,Chen:2020} focused primarily on metallic, superconducting, and insulating phases emerging from interacting domain wall networks.  
These works emphasized how interaction-generated scattering processes in \moire systems can destabilize the quadratic fixed point and lead to competing correlated states, even in the absence of explicit disorder.

A common theme emerging from these studies is that interaction-induced scatterings can generate a rich phase landscape, including correlated insulators, charge- and spin-density-wave states, and superconductivity.  
From the coupled-wire perspective adopted in this review, such phenomena arise from symmetry-allowed nonquadratic operators added on top of the forward-scattering fixed point in Eq.~\eqref{Eq:H0_eqset}.  
Their relevance is governed by the same interaction-dependent scaling dimensions introduced earlier, implying that electrical control of interaction strength directly reshapes the accessible phase diagram.

This viewpoint naturally bridges continuum analyses of twisted materials with the network-based, bosonization-friendly framework, establishing a unified and tunable description of correlated phenomena in nanoscale \moire systems.

\subsection{Quantum anomalous Hall states and chiral edge modes}
\label{SubSec:moire-umklapp}

A central motivation for developing a coupled-wire description of \moire domain wall networks is that it naturally gives access to genuinely topological states of matter.  
In contrast to band-theoretic approaches, topology here emerges from interaction-driven backscattering processes that gap the bulk while leaving protected chiral edge modes, closely following the logic of interacting quantum Hall constructions,\cite{Kane:2002,Teo:2014}  
but realized without Landau levels.
 
Building on this perspective, Ref.~\citenum{Hsu:2023} provided a systematic extension of the network framework to topological phases in \moire systems using a bootstrapping procedure.  
The key conceptual advance was to classify all the scattering operators allowed by symmetries and conservation laws within the domain wall network, and to identify which of them can become renormalization-group relevant in the presence of strong interactions and the long-wavelength \moire periodicity.   

A unified way to describe these possibilities is to introduce a general operator that encodes all symmetry-allowed scattering processes among domain wall modes,
\begin{align}
O_{\{s_{\ell p\sigma}\}} (x) = & \sum_{m}
\prod_{p}
\psi_{R (m+p) \uparrow}^{s_{Rp\uparrow}}(x)
\psi_{L (m+p) \uparrow}^{s_{Lp\uparrow}}(x)  
\psi_{R (m+p) \downarrow}^{s_{Rp\downarrow}}(x)
\psi_{L (m+p) \downarrow}^{s_{Lp\downarrow}}(x) ,
\end{align} 
where the integer set $\{s_{\ell p\sigma}\}$ specifies the number of fermion operators participating in each channel.  
Different choices of $\{s_{\ell p\sigma}\}$ therefore label distinct scattering processes, including intra- and interwire processes as well as higher-order correlated scatterings involving multiple domain walls. 
We note that the branch index is omitted here for simplicity and can be restored straightforwardly.

The central observation is that the allowed sets $\{s_{\ell p\sigma}\}$ are strongly constrained by conservation laws, which enables a systematic bootstrap of all physically admissible scattering operators.  
Energy conservation is ensured by restricting attention to low-temperature processes involving modes near the Fermi level.  
In the absence of proximity-induced pairing, global particle number or charge conservation further imposes the constraint
\begin{eqnarray}
\sum_{p,\sigma} \left(s_{Rp\sigma}+s_{Lp\sigma}\right) = 0 .
\end{eqnarray} 

\begin{figure}[th]
\centerline{\includegraphics[width=0.5\linewidth]{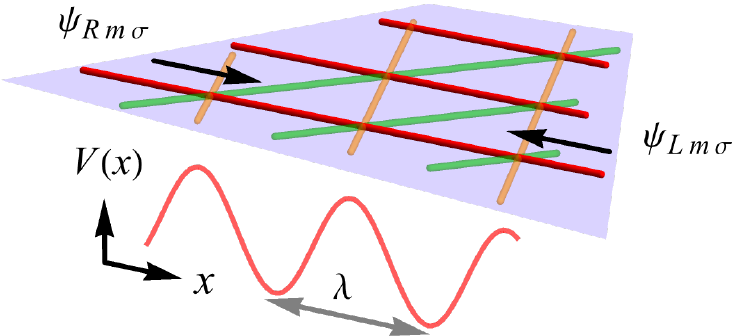}} 
\caption{Schematic illustration of electrons, $\psi_{\ell m \sigma} $ (with $\ell \in \{R,L\} $ and $\sigma \in \{\uparrow, \downarrow \} $), propagating along the domain walls of the twisted bilayer graphene (purple) and experiencing a periodic potential, $V(x)$,  generated by the \moire pattern. 
The colors of the domain walls correspond to those used in Fig.~\ref{Fig:TBG_dw}.  
}
\label{Fig:moire-potential} 
\end{figure}

A central ingredient unique to \moire systems is the presence of a long-wavelength periodic potential with spatial period $\lambda$; see Fig.~\ref{Fig:TBG_dw} and Fig.~\ref{Fig:moire-potential} for illustration.  
Electrons therefore experience an additional \moire crystal momentum, defined modulo the reciprocal lattice vector $2\pi/\lambda$, which relaxes strict momentum conservation.  
As a consequence, generalized umklapp scattering processes become allowed.  
These \moire umklapp terms can become resonant at fractional fillings and gap out the bulk modes of the network, thereby stabilizing correlated insulating states or interaction-driven topological phases.

Combining global charge conservation with generalized momentum conservation yields a resonance condition for the filling factor $\nu = k_F \lambda / \pi$ (for nonzero integer $P$),
\begin{eqnarray}
\nu = \frac{P}{\sum_{p,\sigma} s_{Rp\sigma}}, 
\end{eqnarray}
which specifies the commensurate fillings at which \moire umklapp processes can be allowed.  
At these fillings, interaction-enabled umklapp scattering provides a natural mechanism for opening bulk gaps within the coupled-wire description when they are relevant in the renormalization-group sense, with the possibility of realizing either correlated insulating phases or topological states supporting chiral edge modes.

\begin{figure}[th]
\centerline{\includegraphics[width=\linewidth]{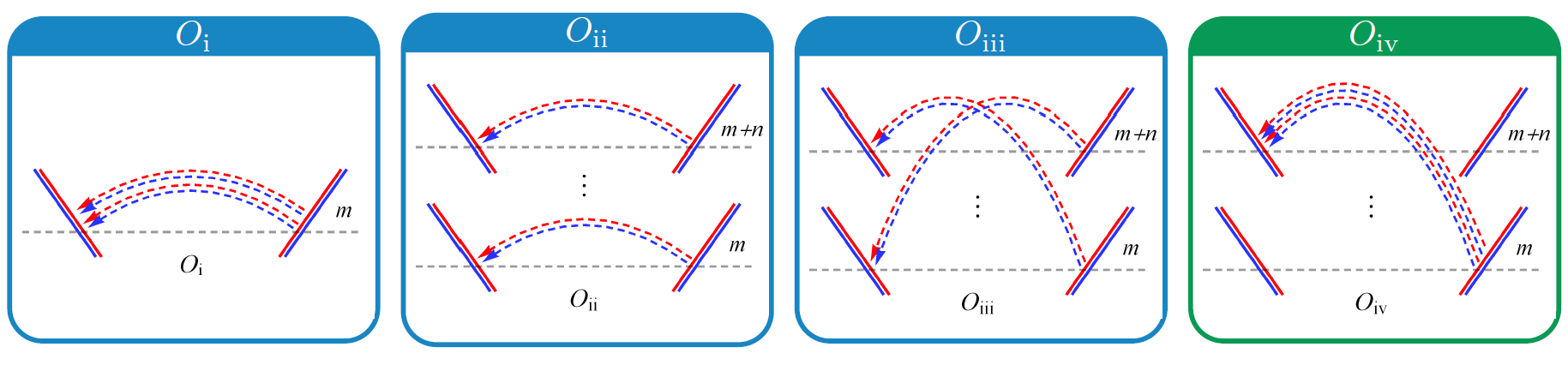}} 
\caption{Categories of \moire umklapp scatterings allowed at the filling factor $\nu = 1/4$: 
intrawire scatterings $O_{\rm i}$ involving a single domain wall;
correlated intrawire scatterings $O_{\rm ii}$ involving multiple domain walls;
interwire scatterings $O_{\rm iii}$ that conserve particle number on each wall; and
interwire scatterings $O_{\rm iv}$ that violate particle-number conservation on individual walls.
Among the four categories,  $O_{\rm iv}$ can lead to a gapped bulk with gapless chiral edge modes. 
}
\label{Fig:moire-umklapp} 
\end{figure}

Within the fermion language, these processes appear as multi-electron backscattering operators whose scaling dimensions are reduced by interactions.  
The bootstrapping procedure based on the particle-number and (generalized) momentum conservation enables a systematic construction of all the allowed \moire umklapp scatterings, summarized schematically in Fig.~\ref{Fig:moire-umklapp}.  
These processes can be organized into four broad categories:
(i) $O_{\rm i}$, intrawire scatterings involving a single domain wall;
(ii) $O_{\rm ii}$, correlated intrawire scatterings involving multiple domain walls;
(iii) $O_{\rm iii}$, interwire scatterings that conserve particle number on each wall; and
(iv) $O_{\rm iv}$, interwire scatterings that violate particle-number conservation on individual walls.
Notably, higher-order scatterings that are typically irrelevant in conventional quantum wires can become decisive in \moire systems, owing to the narrow bandwidth of the domain wall modes.\cite{Wang:2024}

Interestingly, while all ${\rm O}_{\rm i}$, ${\rm O}_{\rm ii}$, ${\rm O}_{\rm iii}$, and ${\rm O}_{\rm iv}$ \moire umklapp processes are capable of driving the system into correlated phases with fractionalized excitations, they lead to qualitatively distinct outcomes.  
In particular, ${\rm O}_{\rm iv}$-type processes generate a fully gapped bulk accompanied by robust chiral edge modes, thereby realizing an interacting version of the quantum anomalous Hall effect within the coupled-wire description.  
Within the network picture, the structure of the renormalization-group relevant backscattering operator directly fixes the number of gapless edge modes, which naturally produces a sequence of quantum anomalous Hall states, or Chern insulators, at fractional fillings.

Beyond reproducing integer and fractional Chern insulating behavior, the same methodology predicts more exotic topological phases.  
The construction in Ref.~\citenum{Hsu:2023} also allows for the possibility of stabilizing electronic states that are fractionalized analogues of the quantum spin Hall effect.  
These phases can be viewed either as fractional quantum spin Hall states or, equivalently, as time-reversal-invariant counterparts of fractional quantum Hall states.  
In the network language, they arise when interaction-induced backscattering gaps the bulk while preserving time-reversal symmetry, leaving counterpropagating fractionalized edge modes. Notably, subsequent experiments in \moire systems
have reported signatures with such fractional quantum spin Hall behavior. Specifically, transport measurements on twisted MoTe$_2$ revealed a state with zero anomalous Hall conductivity together with a fractionally quantized edge conductance per edge, which was interpreted as evidence for a fractional quantum spin Hall insulator.\cite{Kang:2024a} In the same work, integer quantum spin Hall states were also reported at different fillings. These observations highlight \moire transition-metal dichalcogenides as a promising platform for realizing a variety of time-reversal-symmetric topological phases.

\begin{figure}[t]
\centerline{\includegraphics[width=\linewidth]{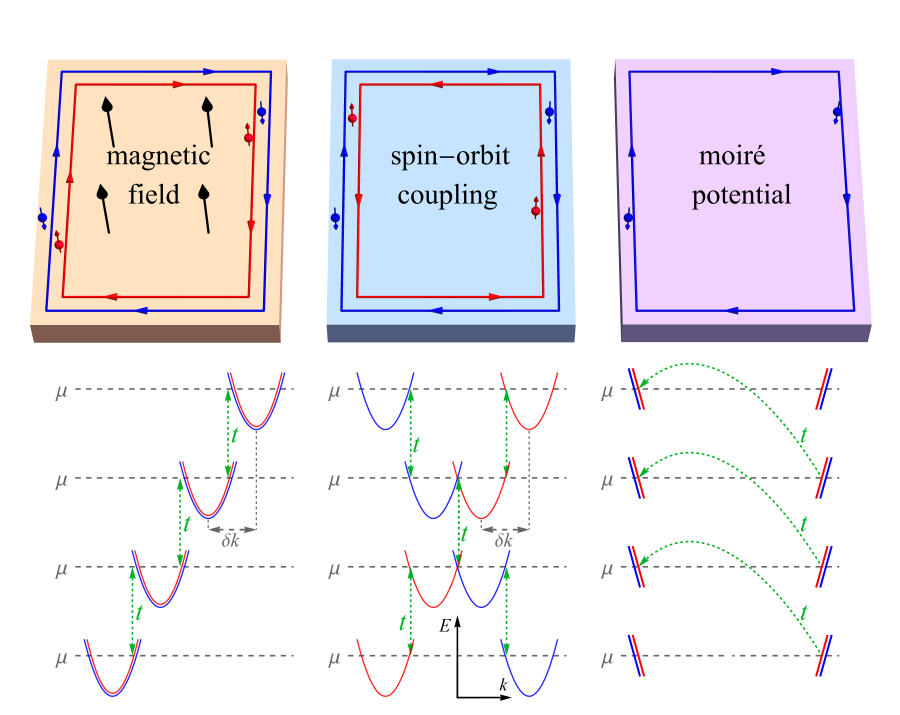}} 
\caption{Trio of quantum Hall phenomena (top) and their corresponding coupled-wire description (bottom).
From left to right: (fractional) quantum Hall effect, with the coupled-wire construction by Ref.~\citenum{Kane:2002},
quantum spin Hall effect with Ref.~\citenum{Klinovaja:2014}, and quantum anomalous Hall effect with Ref.~\citenum{Hsu:2023}.
For the first two systems, a shift $\delta k$ in the band dispersion across wires is induced by external magnetic fields or spin-orbit coupling. In contrast, in quantum anomalous Hall states the momentum mismatch between the initial and final scattering states is compensated by the crystal momentum supplied by the \moire potential.
}
\label{Fig:QH-trio} 
\end{figure}

Drawing on seminal works on coupled-wire constructions of integer and fractional quantum Hall states,\cite{Kane:2002} this line of research establishes a purely interaction-driven route to realizing integer and fractional quantum anomalous Hall states in \moire systems, without invoking external magnetic fields.  Remarkably, the coupled-wire construction now spans the full trio of quantum Hall phenomena.\cite{Kane:2002,Klinovaja:2014,Hsu:2023} As illustrated in Fig.~\ref{Fig:QH-trio}, the coupled-wire framework now provides a unified description encompassing the quantum Hall effect,\cite{Kane:2002} the quantum spin Hall effect,\cite{Klinovaja:2014} and the quantum anomalous Hall effect,\cite{Hsu:2023} highlighting the power of one-dimensional interacting building blocks in generating a broad hierarchy of topological states.

An appealing feature of the coupled-wire  description is that it translates the topological structure of \moire correlated states into experimentally accessible signatures.  
Rather than depending on microscopic interaction details, these signatures are fixed by the universal properties of the effective edge theory, providing a clear diagnostic of interaction-enabled topology.
A primary example is the spectral features in the scanning tunneling spectroscopy, which probes the local density of states at an edge.
For a single fractional edge mode, the density of states obeys a universal scaling form in temperature $T$ and energy $\epsilon$,
\begin{eqnarray}
\rho_{\rm dos,~edge} (\epsilon,T) &\propto&
T^{\frac{1}{f}-1}
\cosh\!\left(\frac{\epsilon}{2k_{\rm B}T}\right)
\left|
\Gamma\!\left(
\frac{1}{2f}
+ i \frac{\epsilon}{2\pi k_{\rm B}T}
\right)
\right|^{2},
\label{Eq:QAHE_edge}
\end{eqnarray}
which reduces to the power law $\rho(\epsilon)\propto |\epsilon|^{1/f-1}$ in the low-temperature limit.
The exponent is fixed solely by the universal fractional parameter $f$ characterizing the edge mode and is independent of the forward-scattering details.
This behavior contrasts sharply with tunneling into domain wall modes in the network interior [see Eq.~\eqref{Eq:DOS}], where scaling exponents depend on the details of the interaction parameters and device geometry.

\begin{figure}[th]
\centerline{
\includegraphics[width=0.3\linewidth]{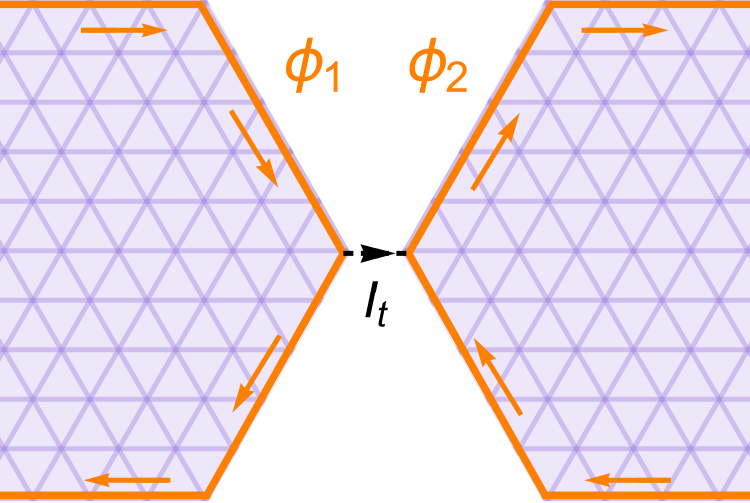} 
\includegraphics[width=0.3\linewidth]{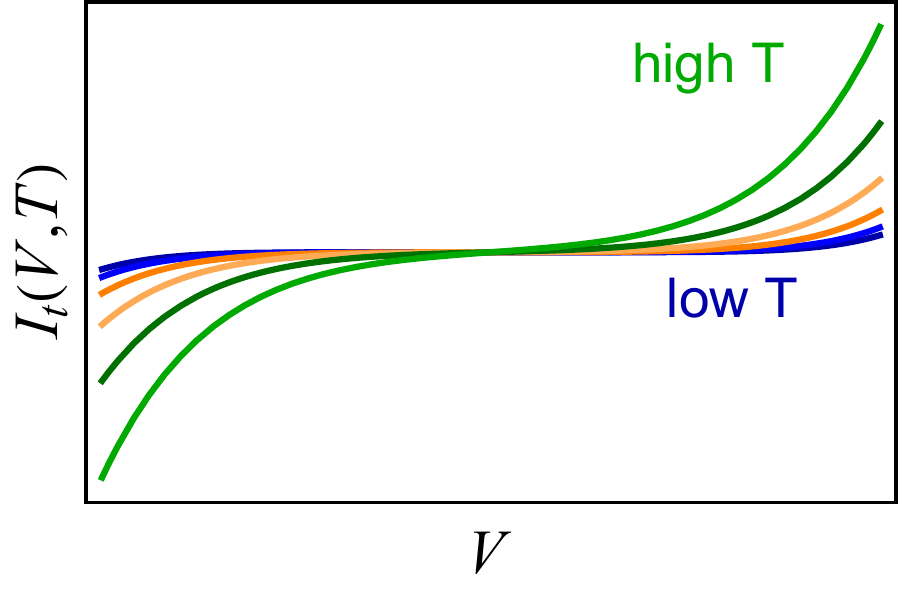} 
\includegraphics[width=0.3\linewidth]{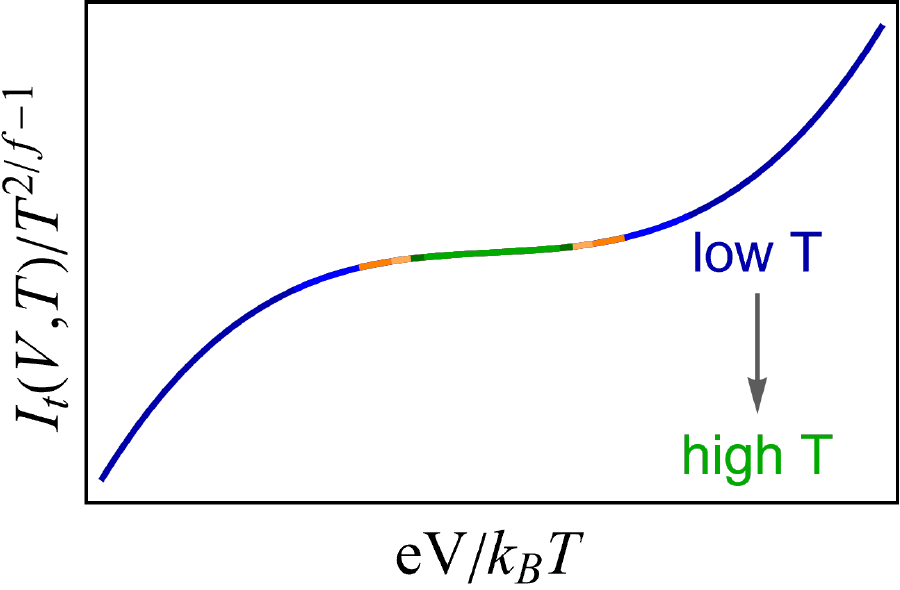}
} 
\caption{Left: Schematic of an interedge tunneling setup between two chiral bosonic  modes $\phi_1$ and $\phi_2$  on the two edges. 
Middle: Representative current-bias characteristics obtained from the tunneling expression in Eq.~\eqref{Eq:It-V}. 
Right: Rescaled current-bias curves demonstrating universal scaling, where data measured at different temperatures collapse onto a single curve. 
}
\label{Fig:QAHE-QPC1} 
\end{figure}

In addition to local spectroscopy, edge transport provides another direct probe of the fractionalized edge modes predicted by the coupled-wire description.  
Because the low-energy degrees of freedom are (fractional) chiral edge fields, transport observables are governed by universal scaling laws fixed by the fractional parameter $f$.

The first setup involves interedge tunneling between two nanoflakes brought into close proximity, as schematically shown in Fig.~\ref{Fig:QAHE-QPC1}. When the system is tuned to a (fractional) quantum anomalous Hall state via the filling factor, evaluating the tunneling current between chiral edge fields gives the current-bias relation at temperature $T$,
\begin{eqnarray}
I_{\rm t} &\propto&
T^{\frac{2}{f}-1}
\sinh\!\left(\frac{eV}{2k_{\rm B}T}\right)
\left|
\Gamma\!\left(
\frac{1}{f}
+ i \frac{eV}{2\pi k_{\rm B}T}
\right)
\right|^{2},
\label{Eq:It-V}
\end{eqnarray}
where $V$ is the applied bias.  
This expression defines a universal scaling function controlled solely by $f$, independent of microscopic edge details.

\begin{figure}[th]
\centerline{
\includegraphics[width=0.35\linewidth]{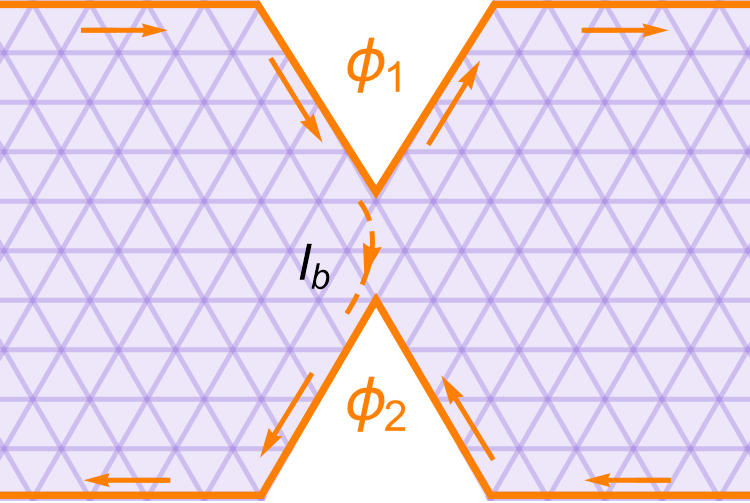}
\includegraphics[width=0.35\linewidth]{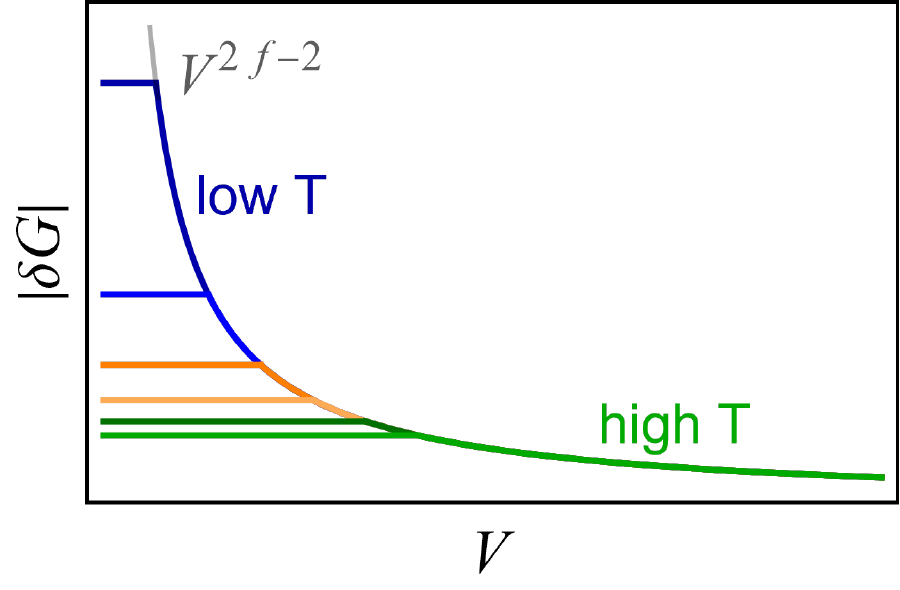}
} 
\caption{Left: Schematic illustration of interedge backscattering between two chiral edge modes; $\phi_1$ and $\phi_2$ denote the chiral bosonic fields on the two edges, and $I_b$ is the backscattering current. 
Right: Differential conductance correction $\delta G$ to the quantized edge conductance arising from interedge backscattering with a nonuniversal backscattering strength.
}
\label{Fig:QAHE-QPC2} 
\end{figure}

A complementary setup employs a quantum point contact, where interedge backscattering is weak but finite, as illustrated in Fig.~\ref{Fig:QAHE-QPC2}.  
In this case, the leading observable is the correction $\delta G$ to the quantized edge conductance.  
At low energies, one finds the power-law behavior
\begin{eqnarray}
\label{Eq:edge-power-law}
|\delta G| &\propto&
\begin{cases}
V^{2f-2}, & eV \gg k_{\rm B}T, \\
T^{2f-2}, & eV \ll k_{\rm B}T ,
\end{cases}
\end{eqnarray}
again with the exponents fixed by the fractional parameter $f$.

Together, interedge tunneling and backscattering measurements provide transport diagnostics that are directly sensitive to fractionalization and topology. 
Their universal scaling forms clearly distinguish topological \moire correlated states from the nonuniversal transport behavior associated with domain wall modes in the network interior. 
The contrast between the edge scaling and the bulk domain wall scaling, observable in both spectroscopy and transport, therefore offers a clean and experimentally accessible criterion for identifying topological \moire correlated states within the coupled-wire description.

\section{Beyond purely electronic systems and other platforms}
\label{Sec:BeyondTBG}

The coupled-wire perspective developed above is not limited to twisted bilayer graphene.
In addition to purely electronic systems, one can introduce local magnetic moments and explore the interplay between the local moments and itinerant electrons in the domain walls. 
More broadly, one-dimensional electronic channels can emerge in a variety of \moire  layered systems, providing additional platforms where interaction-driven physics can be explored within a unified framework.  
For example, in chiral twisted trilayer graphene, one-dimensional modes emerge even in the absence of an interlayer bias.\cite{Devakul:2023a, Nakatsuji:2023}  
Beyond graphene-based materials, related domain wall or interface channels have also been identified in other van der Waals systems, such as twisted bilayer WTe$_2$,\cite{Wang:2022,Yu:2023} where strong spin-orbit coupling and band topology further enrich the low-energy behavior.

\subsection{Interplay between itinerant carriers and localized moments}
\label{SubSec:2Dhelix}

\begin{figure}[th]
\centerline{
\includegraphics[width=0.35\linewidth]{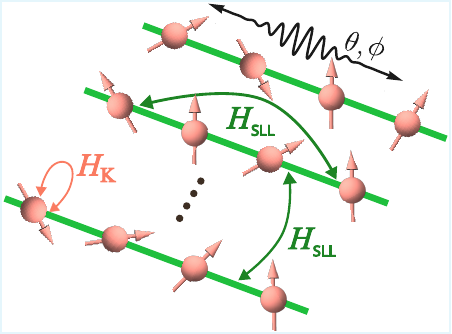} 
\includegraphics[width=0.6\linewidth]{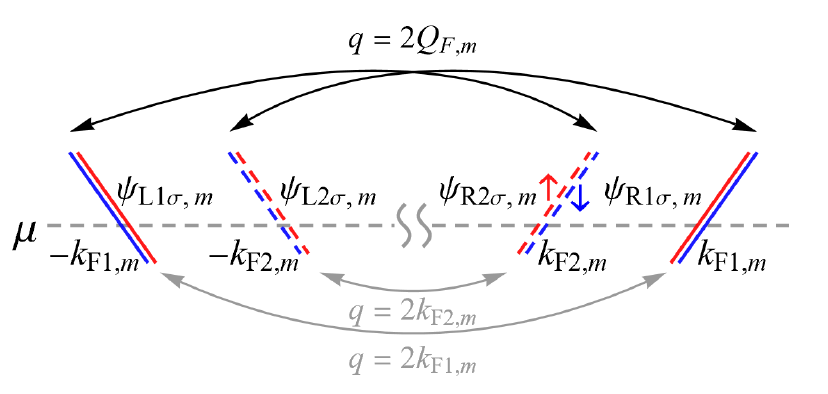} 
} 
\caption{Left: Schematic of a system comprising localized moments (pink) and sliding Luttinger liquids (green), and the Kondo coupling between the two subsystems.
Right: In the domain wall spectrum, electrons backscatter between Fermi points. Due to the multiple energy branches, there arise $2Q_F$, $2k_{F1}$, and $2k_{F2}$-scatterings. 
}
\label{Fig:SLL} 
\end{figure}

Beyond hosting purely electronic correlated phases, domain wall networks also provide a natural platform to revisit a classic theme in condensed matter physics: the interplay between itinerant carriers and localized spins. 
Here we consider systems in which interacting electrons propagating along domain walls couple to localized magnetic moments through a Kondo-type exchange interaction, as schematically illustrated in Fig.~\ref{Fig:SLL}. 
The electronic subsystem realizes a sliding Luttinger liquid along the domain wall direction,\cite{Chen:2020,Hsu:2023} while the localized moments can be engineered either via magnetic adatoms deposited on graphene~\cite{Hong:2012} or via hyperfine-coupled nuclear spins using isotope control.\cite{Fischer:2009,Churchill:2009} 
This Kondo-sliding Luttinger liquid setting extends well beyond the extensively studied strictly one-dimensional cases and higher-dimensional noninteracting systems, and crucially leverages the electrical tunability intrinsic to nanoscale \moire structures.

The full system is described by
\begin{align}
H_{\rm tot} = H_{\rm SLL} + H_{\rm K}, \qquad 
H_{\rm K} = J_K \sum_j \mathbf{S}_j \cdot \mathbf{s}(\mathbf{x}_j),
\end{align}
where $H_{\rm SLL}$ captures the interacting itinerant carriers confined to the domain walls, and $H_{\rm K}$ describes their exchange coupling to localized (classical) spins $\mathbf{S}_j$ at positions $\mathbf{x}_j$.  As before, the electronic subsystem is conveniently formulated in bosonized language as a sliding Luttinger liquid,
\begin{align}
H_{{\rm SLL}}  = & \frac{1}{N_{\rm dw}} \sum_{q_{\perp}}
\int\frac{\hbar dx }{2\pi}
\left[
\frac{v_{cS}(q_{\perp})}{ K_{cS}(q_{\perp})}
\left|\partial_{x }\phi_{cS,q_{\perp}}(x)\right|^{2}
+ v_{cS}(q_{\perp}) K_{ cS }(q_{\perp})
\left|\partial_{x}\theta_{cS,q_{\perp}} (x)\right|^{2}
\right] \nonumber  \\
& + H_{cA} + H_{sS} + H_{sA}. 
\label{Eq:H_SLL}
\end{align}
As introduced for Eq.~\eqref{Eq:H0_original} we have an interacting, charge-symmetric sector, while the other sectors are effectively noninteracting and given by Eq.~\eqref{Eq:H0s_original}. 
In contrast to Eq.~\eqref{Eq:H0cS_original}, however, here we assume periodic boundary conditions in the direction perpendicular to the domain walls, which permits the introduction of a transverse momentum $q_{\perp}$.  
Within this formulation, the interaction dependence of the electronic fixed point enters through the $q_{\perp}$-dependent parameters $K_{cS}(q_{\perp})$ and $v_{cS}(q_{\perp})$, which encode how Coulomb interactions correlate parallel domain walls and are conveniently represented by a Fourier expansion of $K_{cS}(q_{\perp})$.  
Although this assumption can be relaxed and treated numerically, as demonstrated in Sec.~\ref{Sec:network}, the present approach follows earlier sliding Luttinger liquid constructions,\cite{Emery:2000,Vishwanath:2001,Mukhopadhyay:2001a,Mukhopadhyay:2001b,Chen:2020,Hsu:2023} and is sufficient to capture the essential physics of the Kondo problem in coupled-wire systems.  
Compared with the original proposals for cuprates,\cite{Emery:2000,Vishwanath:2001,Mukhopadhyay:2001a,Mukhopadhyay:2001b} a key advantage of \moire domain wall networks is that the interaction strength can be estimated directly from continuum modeling and tuned electrically, as discussed in Sec.~\ref{SubSec:quadratic}.

The coupling between itinerant electrons  and localized moments is described by a Kondo-type  interaction,
\begin{align}
   H_{{\rm K}} = \sum_{j,m}\sum_{\mu , \sigma , \sigma^{\prime}} 
    J_K^{\mu}  
   \left[ 
   \psi^{\dagger} _{\sigma,m}( x_{j}) \sigma^\mu_{\sigma\sigma^{\prime}}
   \psi_{\sigma^\prime,m}( x_{j}) 
   \right]   
   S^\mu_{m} ( x_{j}),
   \label{Eq:H_K}
\end{align}
which locally couples the spin density of the electrons to localized magnetic moments, with the effective coupling strength $ J_K^{\mu} $ and   $\mu\in\{x,y,z\}$.

To proceed, we focus on the weak-coupling regime in which the Kondo temperature is parametrically smaller than all the other relevant electronic energy scales.  
In this limit, the dominant role of the exchange term $H_{\rm K}$ is the generation of an indirect Ruderman-Kittel-Kasuya-Yosida (RKKY) interaction between localized moments mediated by the correlated  itinerant electrons.  
Formally, integrating out the electronic degrees of freedom to the second order in $J_K$ leads to an effective interaction of the form\cite{Simon:2008,Braunecker:2009a,Braunecker:2009b}
\begin{align}
    H_{\rm R} = \sum_{m,n}\sum_{\mu}
    \int  dr\,dr^\prime 
    J_{n}^{\mu}(r-r^\prime)
    S^{\mu}_{m+n}(r)\,S_m^{\mu}(r^\prime),
    \label{Eq:RKKY}
\end{align}
where the exchange kernel $J_{n}^{\mu}(r)$ is spatially oscillatory and proportional to the spin susceptibility of the interacting domain wall modes.  
This RKKY interaction couples localized moments both within a given domain wall and across parallel domain walls, providing the microscopic origin of collective magnetic ordering in the coupled wires.

The momentum-space structure of $J_{n}^{\mu}$ directly reflects the dominant backscattering processes of the domain wall modes indicated in Fig.~\ref{Fig:SLL}.  
In addition to intrabranch processes with momentum transfer $q=\pm 2k_{F1}$ or $\pm 2k_{F2}$, the interbranch processes with
$q=\pm 2Q_F\equiv \pm(k_{F1}+k_{F2})$ play a more significant role
due to a higher number of states available for scatterings.  
As a result, the electronic spin susceptibility exhibits pronounced features near $q\simeq \pm 2Q_F$, favoring an oscillatory RKKY interaction at this wave vector.  
The energy is therefore minimized when the localized moments form a helical texture along the domain walls with pitch
$\lambda_{\rm hx}\approx \pi/Q_F$.

This helical order is a natural analogue of the RKKY-induced spin helix in purely one-dimensional systems,\cite{Braunecker:2009a,Braunecker:2009b,Meng:2013,Klinovaja:2013a,Meng:2014a,Hsu:2015,Hsu:2017,Hsu:2018b} but with an important distinction arising from the coupled-wire geometry.  
The presence of nonlocal interwire couplings $J_{n\neq 0}^{\mu}$ in Eq.~\eqref{Eq:RKKY} energetically favors a uniform relative phase between helices on neighboring parallel domain walls, locking their offset phases.  
This interwire phase locking generates spatial coherence across parallel domain walls and gives rise to an emergent two-dimensional spin helix.  
Compared with isolated channels, the resulting collective order substantially enhances the ordering temperature.\cite{Chang:2025}  
Moreover, the spatially rotating exchange field associated with the helix partially gaps the electronic spectrum, producing a Peierls-like energy gain that further stabilizes the ordered phase.

\begin{figure}[th] 
\centerline{ 
\includegraphics[width=0.42\linewidth]{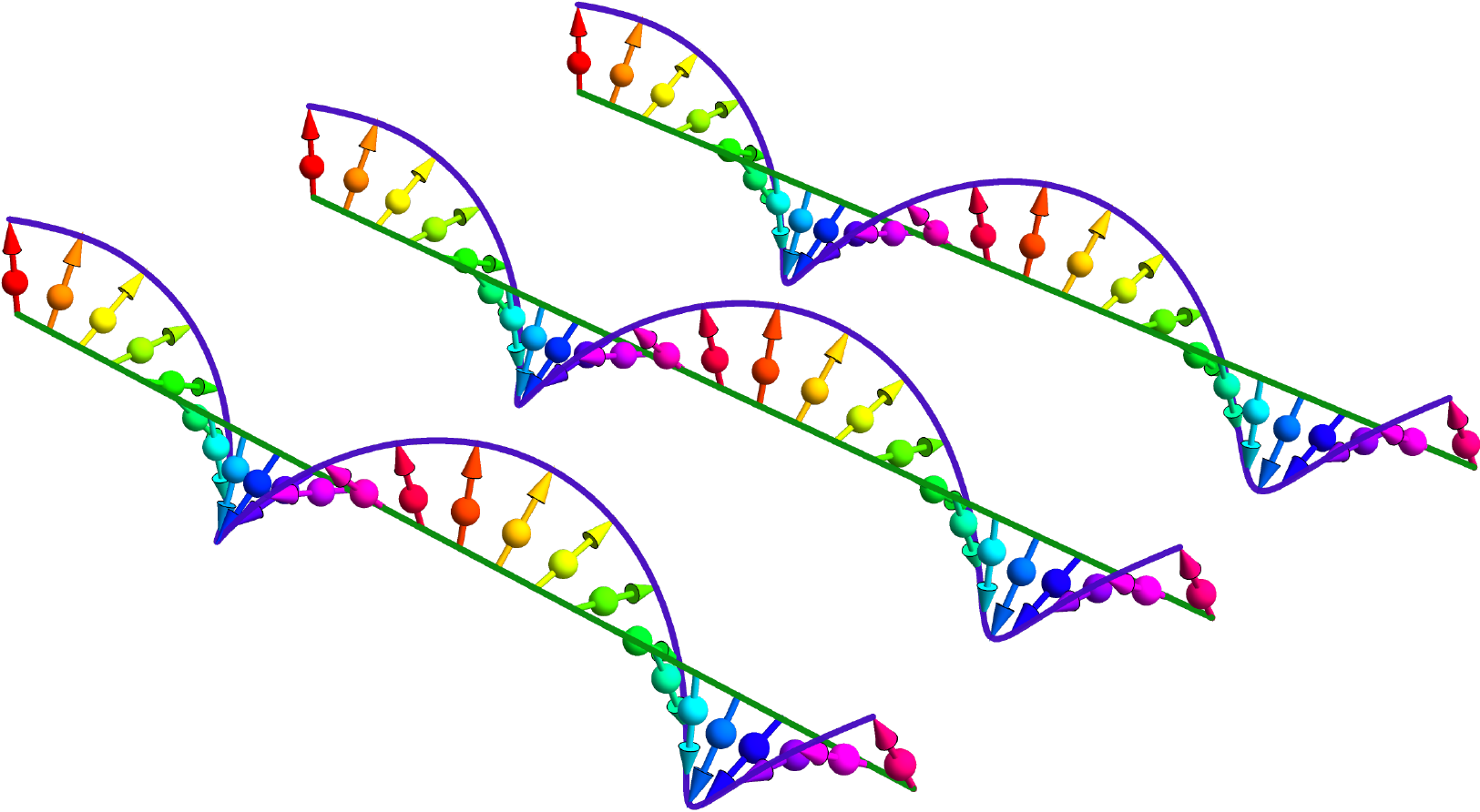} 
\includegraphics[width=0.245\linewidth]{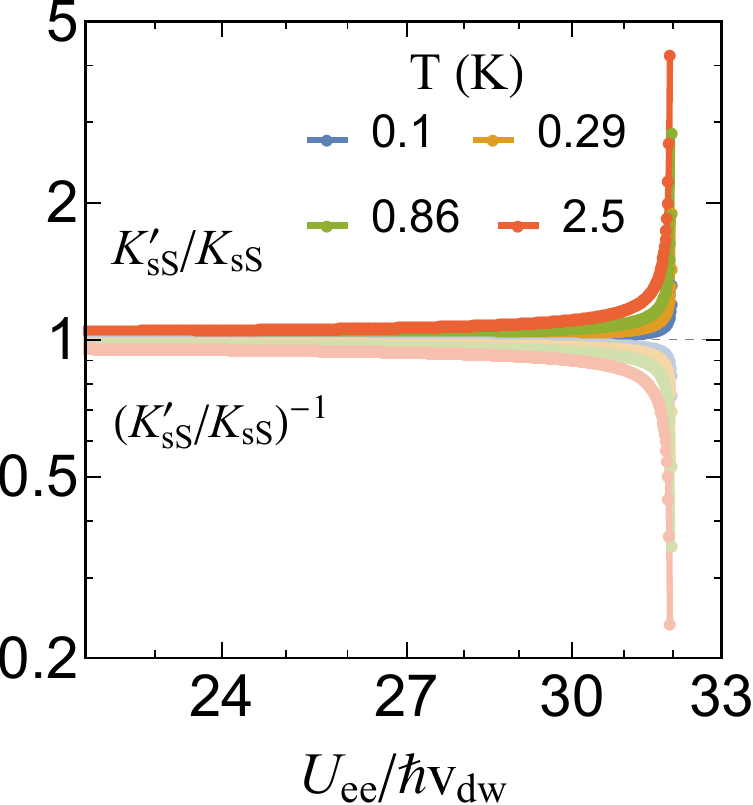} 
\includegraphics[width=0.26\linewidth]{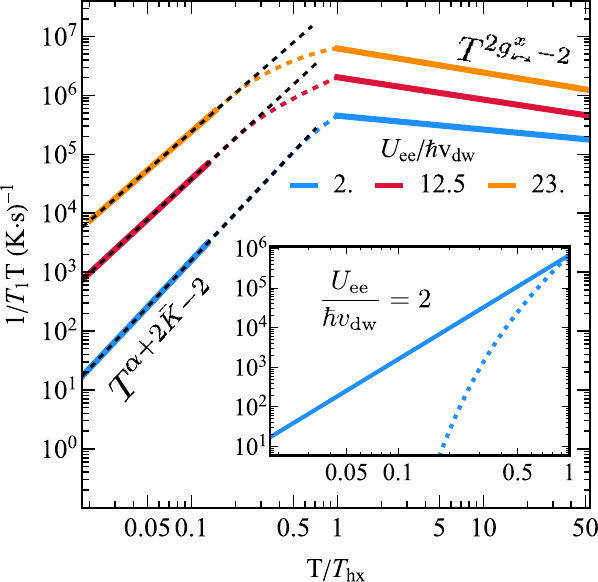} 
} 
\caption{Left: illustration of the spatially phase coherent spin helix stabilized by the RKKY interaction in a coupled-wire system.
Middle: Scaling exponent as a function of the electron interaction strength $(U_{\rm ee})$.
Right: Spin relaxation rate $1/T_1$ as a function of temperature, exhibiting two distinct power-law regimes above and well below $T_{\rm hx}$. The right two panels are reproduced from Ref.~\citenum{Chang:2025}. 
} 
\label{Fig:2D-helix} 
\end{figure}

Formally, as displayed in Fig.~\ref{Fig:2D-helix}, the localized moments develop a helical configuration of the form
\begin{equation}
\mathbf{S}( x) =S_0
\left[
\cos(2Q_F x +\phi_0)\,\hat{\mathbf{e}}_1
+\sin(2Q_F x +\phi_0)\,\hat{\mathbf{e}}_2
\right],
\end{equation}
where $\hat{\mathbf{e}}_{1,2}$ are orthogonal unit vectors perpendicular to the domain wall direction.  
Overall, the RKKY mechanism in the coupled-wire network illustrates how one-dimensional electron correlations qualitatively reshape itinerant-spin physics, leading to collective magnetic order without a direct analogue in strictly one-dimensional systems.

Beyond establishing the tendency toward a two-dimensional spin helix, it is natural to examine the low-energy collective excitations of the ordered phase and their feedback on the electronic subsystem.  
At finite temperatures, the relevant collective modes are magnons, whose lowest excitation energy is set by the RKKY scale, as shown by standard spin-wave analysis.\cite{Chang:2025}  
A key advance is the recognition that these collective modes can couple back to the itinerant electrons in a nontrivial manner.  
Ref.~\citenum{Chang:2025} analyzed this feedback by considering the coupling between magnons and the spin degrees of freedom of the domain wall electrons.

The central qualitative message is that magnons influence the electronic subsystem not only through backscattering processes associated with spin flips, but also through forward-scattering channels that renormalize scaling dimensions.  
This interaction is the spin-sector analogue of the electron-phonon coupling discussed in Sec.~\ref{Sec:Phonon-in-TBG}, with magnons playing the role of the collective bosonic modes generated by the electronic correlations themselves.
 
Retaining the quadratic form, the coupled electron-magnon system can be diagonalized nonperturbatively, leading to a renormalized parameter $K_{sS}$.  The most striking consequence is the emergence of a magnon-induced singularity, as shown in Fig.~\ref{Fig:2D-helix}, where the scaling dimensions of operators constructed from $\phi_{sS}$ ($\theta_{sS}$) diverge (vanish), signaling a breakdown of conventional low-energy scaling behavior. The mechanism closely parallels the forward-scattering singularities familiar from electron-phonon coupled one-dimensional systems,\cite{Bardeen:1951,Wentzel:1951,Engelsberg:1964,
Loss:1994,Martin:1995,Hsu:2024}
generalized for coupled-wire systems,\cite{Wang:2024} as discussed in Sec.~\ref{SubSec:quadratic}. 
Crucially, this singularity is experimentally accessible because the underlying electron-electron interaction strength can be tuned via the interlayer bias and electrostatic screening.\cite{Wang:2024}  

From an experimental perspective, direct real-space imaging of the spin helix may be possible using spin-resolved probes, although the presence of metallic gates in typical devices can limit such approaches.  
More generally applicable diagnostics are therefore provided by transport and magnetic response measurements.  
In mesoscopic transport, the partial gap opened by the spin helix is expected to reduce the conductance below the ordering temperature $T_{\rm hx}$,\cite{Braunecker:2009a,Braunecker:2009b,Hsu:2015,Hsu:2020} providing an indirect but robust signature of helix formation.

Magnetic resonance measurements offer a complementary probe.  
In particular, the spin relaxation rate $1/T_1$ exhibits two distinct power-law regimes above and well below $T_{\rm hx}$,\cite{Chang:2025}  reflecting the crossover from the sliding Luttinger liquid regime to the helix-ordered phase, as illustrated in Fig.~\ref{Fig:2D-helix}.  
The corresponding exponents are controlled by interaction-dependent parameters of the network, generalizing the Korringa law and directly revealing the correlated nature of the electronic subsystem.\cite{Korringa:1950,Coleman:2015}  
Notably, as the system approaches the magnon-induced singularity, the low-temperature relaxation rate is predicted to diverge, providing a sharp and experimentally accessible signature of this singular regime.

By integrating Luttinger liquid physics with Kondo-type interactions, the domain wall network thus supports a two-dimensional spin helix and a magnon-induced singularity whose properties can be tuned by electric fields or the twist angle.\cite{Chang:2025}   
These phenomena deepen our understanding of strongly correlated electron systems and highlight twisted van der Waals structures as electrically controllable platforms for exploring collective magnetic order and emergent singular behavior.

\subsection{Other platforms}
\label{SubSec:Others}

Although much of the discussion above is motivated by twisted bilayer graphene, the underlying ideas are considerably more general.  
Similar one-dimensional channels have been identified or proposed in a variety of nanoscale systems,\cite{Kennes:2020,Fujimoto:2022,Wang:2022,Devakul:2023a,Du:2023,Nakatsuji:2023,Yu:2023,Klein:2023} where spatial modulation, topology, or symmetry breaking confines low-energy electronic modes to narrow regions.

Prominent examples include chiral twisted trilayer graphene,\cite{Devakul:2023a,Nakatsuji:2023} where symmetry and stacking chirality generate extended one-dimensional channels even in the absence of an interlayer bias, and twisted bilayers of WTe$_2$,\cite{Wang:2022,Yu:2023,Wu:2024prl,Hu:2024} where strong spin-orbit coupling and intrinsic topology provide additional control over the resulting network.  
Notably, twisted bilayer WTe$_2$ can realize two distinct types of coupled-wire systems within the same platform, depending on the carrier type.\cite{Wu:2024prl} On the electron-doped side, the reconstructed \moire channels host four-flavor modes on each wire, similar to those discussed for twisted bilayer graphene.\cite{Wang:2024} By contrast, on the hole-doped side, the system realizes a more conventional two-flavor coupled-wire structure. 
In these systems, the basic ingredients emphasized in this review, geometrically confined one-dimensional modes, electrically tunable interactions, and interchannel coupling, are naturally present, making them promising platforms for correlated and topological phases.

More broadly, the \moirenospace-based coupled-wire mechanism can, in principle, operate in a wide class of engineered nanoscale settings, including engineered topological insulators\cite{Fujimoto:2022} and strain-defined networks in graphene.\cite{Hsu:2020n}  
Another realization involves helical liquids in the edge channels of quantum spin Hall insulators,\cite{Hsu:2021,Hsu:2025} where stacked structures have been proposed as a platform for realizing sliding helical liquids,\cite{Hung:2025b} representing a helical analogue of the sliding Luttinger liquid. 
In all these cases, arrays of one-dimensional channels provide a natural starting point for interaction-driven phenomena that go beyond single-wire physics.

\begin{table}[th]
\caption{Summary of representative electronic phases in interacting \moire domain wall networks, their key parameters, and main experimental signatures. 
}
\begin{minipage}{\textwidth}
\begin{tabular}[c]{ p{3.2cm} p{3.4cm} p{5.5cm} }
\hline \hline
\centering
Phase & Key parameters & Main experimental signatures \\
\hline
Correlated metal\footnote{See Refs.~\citenum{Xu:2019,Chou:2019,Chen:2020,Wang:2024,Wu:2024prl,Hu:2024}.} & 
Interaction strength.\footnote{In particular, the electron-electron interaction strength in the charge-symmetric sector, $U_{\phi_{cS},n}$, can be tuned through the interlayer bias $V_d$, screening length $d$, dielectric constant $\varepsilon_r$, and twist angle $\theta$.
} 
  & Universal scaling in scanning tunneling spectroscopy [see Eq.~\eqref{Eq:DOS}]; power-law dependencies on temperature and bias in transport measurement. \\
 
 \hline
Density waves
\newline 
  (CDW/SDW)\footnote{See Refs.~\citenum{Chou:2019,Chen:2020,Wang:2024,Wu:2024prl}.} 
& Interaction strength. &  Gapped domain wall spectrum; suppression of conductance at low temperatures. \\

 \hline
Superconductivity\footnote{See Refs.~\citenum{Xu:2019,Chou:2019,Chen:2020,Wang:2024,Wu:2024prl}.}  & Interaction strength, 
\newline
electron-phonon coupling.
& Zero resistance and suppression of the local density of states at low temperatures. \\

 \hline
Anderson insulator\footnote{See Ref.~\citenum{Wang:2024}.} & Interaction strength, 
\newline
disorder strength.  & Metal-insulator transition; tunable localization strength  and temperature.\\

\hline  
Quantum anomalous  Hall states\footnote{See Ref.~\citenum{Hsu:2023}.} & Interaction strength, 
\newline
filling factor.  & Gapped bulk spectrum with gapless edge modes; universal scaling in scanning tunneling spectroscopy at edges [Eq.~\eqref{Eq:QAHE_edge}]; 
universal scaling in interedge tunneling $I$-$V$ curves [see Eq.~\eqref{Eq:It-V}];
power-law conductance corrections in quantum point contacts [see Eq.~\eqref{Eq:edge-power-law}] \\

 \hline
Spin helix\footnote{See Ref.~\citenum{Chang:2025}; this phase requires intentionally induced localized moments, such as magnetic adatoms.} & Interaction strength, 
\newline Kondo exchange coupling. & Conductance reduction below ordering temperature; fractional power law in spin relaxation rate ($1/T_1$).\\

\hline \hline
\end{tabular}
\label{Tab:roadmap}
\end{minipage}
\end{table}

\section{Outlook}
\label{Sec:Outlook}

The coupled-wire perspective on \moire domain wall networks opens several promising directions for future experimental and theoretical investigation.  
On the experimental side, many of the phenomena discussed in this review are directly accessible with existing probes; Table~\ref{Tab:roadmap} summarizes the primary correlated and topological phases discussed in this review, alongside their key tuning parameters and main experimental signatures.
Transport and scanning tunneling spectroscopy can be used to measure the predicted power-law behavior of conductance and local density of states, as well as the contrast between bulk domain wall responses and edge signatures of topological phases.  
In particular, the ability to tune interaction strength \emph{in situ} via interlayer bias and electrostatic screening makes it feasible to explore metal-insulator crossovers driven by Anderson localization within a single device and a single disorder realization.

At the same time, it is useful to distinguish between predictions that are already closely connected to existing experimental probes and those that remain more prospective.  
The existence of reduced-dimensional \moire channel structures is already supported experimentally in more than one platform.  
In small-angle twisted bilayer graphene, scanning-probe and transport experiments have resolved the reconstruction into stacking domains separated by domain walls and have probed the associated one-dimensional channels.\cite{Rickhaus:2018,Xu:2019,Yoo:2019,Verbakel:2021,Ouyang:2026} 
In twisted bilayer WTe$_2$, the corresponding electronic channels have been connected more directly to correlated transport behavior, including strong in-plane anisotropy and power-law conductance consistent with Luttinger-liquid-like physics.\cite{Wang:2022,Yu:2023}  
By contrast, more delicate regimes, including fractional topological phases and disorder-driven crossover phenomena within coupled domain walls, will likely require cleaner devices, improved control of screening and displacement fields, and more refined edge- or channel-selective measurements.  
In this sense, current experiments already access several central ingredients of the coupled-wire description, while the broader landscape of strongly correlated and topological network phases remains an important direction for future work through the transport and spectroscopic probes reviewed here and summarized in Table~\ref{Tab:roadmap}. 

An additional direction concerns the deliberate introduction of localized magnetic moments, for example through magnetic adatoms or isotope engineering.  
While this adds complexity to the system, it also enables controlled realizations of Kondo physics in coupled-wire settings, where the interplay between itinerant electrons, collective magnetic order, and reduced dimensionality can be explored in regimes inaccessible in conventional materials.
While the above analysis focuses on classical spins and the RKKY regime, one can also consider quantum localized spins coupled to the electron spins in the domain wall network, which could be used to explore multichannel Kondo physics.~\cite{Affleck:1993,Affleck:1994,Affleck:2010,Andrei:1984,Parcollet:1998,Kirchner:2020,Wang:2022c}

Another important ingredient that has not been addressed in detail here is the superconducting proximity effect.  
Proximity coupling domain wall networks to superconductors provides a natural route to engineer pairing at junctions and interfaces, and may enable new tests of the coupled-wire descriptions through Josephson physics, Andreev processes, and the emergence of topological superconducting phases in network geometries. 
Notably, the spin helix discussed above can act as a synthetic spin-orbit field. In one dimension,\cite{Klinovaja:2013a,Hsu:2015} it is known to stabilize Majorana zero modes when the system is proximitized by a superconductor. Extending this mechanism to the present setting is nontrivial and will be interesting for future work.

Taken together, these directions highlight the coupled-wire description as a unifying framework in which topology, strong correlations, and nanoscale engineering meet.  
Rather than being restricted to a specific material or twist geometry, this approach offers a controlled route to fractionalized and symmetry-protected topological phases beyond conventional band theory, and is likely to play an important role across a broad range of emerging \moire and nanoscale quantum materials.

\section*{Acknowledgments}

We thank Y.-Y.~Chang, C.-H.~Chung, Y.~Fukusumi, T.~Grover, G.~Möller, M.~Oshikawa, S.~Slizovskiy, T.~Tohyama, and J.~{\^I}-j.~Wang for interesting discussions. 
This work was financially supported by the National Science and Technology Council (NSTC), Taiwan, through Grant No.~NSTC-114-2112-M-001-057, and Academia Sinica (AS), Taiwan through Grant No.~AS-iMATE-114-12.   
C.-H.~H. acknowledges the Yukawa Institute for Theoretical Physics at Kyoto University, where this work was initiated during the long-term workshop ``Progress of Theoretical Bootstrap.''

\section*{ORCID}

\noindent Chen-Hsuan Hsu \url{https://orcid.org/0000-0003-0963-538X}

\noindent Anna Ohorodnyk \url{https://orcid.org/0009-0000-7910-4918}


\end{document}